%% file: main.tex
\PassOptionsToPackage{table}{xcolor}
\documentclass[sigconf]{acmart}

\AtBeginDocument{%
  }

\usepackage{enumitem}

\usepackage{booktabs,tabularx,array}
\usepackage[table]{xcolor} % enables \columncolor in tables
\newcolumntype{Y}{>{\arraybackslash}X}
\newcolumntype{G}{>{\centering\arraybackslash\columncolor{gray!15}}X} % light gray col

\newcolumntype{C}[1]{>{\centering\arraybackslash}m{#1}} % centered, fixed-width

\usepackage{xcolor}
\definecolor{lightpurple}{HTML}{D8BFD8}
\definecolor{lightorange}{HTML}{EED594} 
\definecolor{lightcyan}{HTML}{8AE2D8}
\definecolor{lightblue}{HTML}{CAECFF}

\definecolor{LikertSA}{HTML}{66C2A5} % Strongly agree
\definecolor{LikertAG}{HTML}{B2D4C1} % Agree
\definecolor{LikertNE}{HTML}{D3D3D3} % Neither
\definecolor{LikertDI}{HTML}{E8D68E} % Disagree
\definecolor{LikertSD}{HTML}{FFD92F} % Strongly disagree

\newcommand{\Likert}[1]{%
  \ifnum#1=5 \cellcolor{LikertSA!80}\fi%
  \ifnum#1=4 \cellcolor{LikertAG}\fi%
  \ifnum#1=3 \cellcolor{LikertNE}\fi%
  \ifnum#1=2 \cellcolor{LikertDI}\fi%
  \ifnum#1=1 \cellcolor{LikertSD}\fi%
  #1%
}

\usepackage{makecell}
\usepackage{multirow}

\newcommand{\llmplaceholder}[1]{\texttt{\{#1\}}}

\usepackage{framed}
\newenvironment{llmscriptblock}[1]{%
    \medskip
    \vspace{-0.5em}
    \begin{snugshade*}
    \small\ttfamily 
}{%
    \end{snugshade*}
    \medskip
}
\definecolor{promptgray}{gray}{0.95}
\colorlet{shadecolor}{promptgray}

\copyrightyear{2026}
\acmYear{2026}
\setcopyright{cc}
\setcctype{by}
\acmConference[IUI '26]{31st International Conference on Intelligent User Interfaces}{March 23--26, 2026}{Paphos, Cyprus}
\acmBooktitle{31st International Conference on Intelligent User Interfaces (IUI '26), March 23--26, 2026, Paphos, Cyprus}
\acmPrice{}
\acmDOI{10.1145/3742413.3789083}
\acmISBN{979-8-4007-1984-4/2026/03}

\begin{document}

\title{ReUseIt: Synthesizing Reusable AI Agent Workflows for Web Automation}

\author{Yimeng Liu}
\authornote{This work was done during an internship at Microsoft Research.}
\email{yimengliu@ucsb.edu}
\affiliation{%
  \institution{University of California, Santa Barbara}
  \city{Santa Barbara}
  \state{California}
  \country{USA}
}

\author{Misha Sra}
\email{sra@ucsb.edu}
\affiliation{%
  \institution{University of California, Santa Barbara}
  \city{Santa Barbara}
  \state{California}
  \country{USA}
}

\author{Jeevana Priya Inala}
\email{jinala@microsoft.com}
\affiliation{%
  \institution{Microsoft Research}
  \city{Redmond}
  \state{Washington}
  \country{USA}
}

\author{Chenglong Wang}
\email{chenwang@microsoft.com}
\affiliation{%
  \institution{Microsoft Research}
  \city{Redmond}
  \state{Washington}
  \country{USA}
}

\renewcommand{\shortauthors}{Liu et al.}

\newcommand{\tool}{\textsc{ReUseIt}\,}

\input{sections/0-abstract}

\begin{CCSXML}
<ccs2012>
   <concept>
       <concept_id>10003120.10003121</concept_id>
       <concept_desc>Human-centered computing~Human computer interaction (HCI)</concept_desc>
       <concept_significance>500</concept_significance>
       </concept>
 </ccs2012>
\end{CCSXML}

\ccsdesc[500]{Human-centered computing~Human computer interaction (HCI)}

\keywords{AI agent reusability and interpretability, workflow synthesis}

\begin{teaserfigure}
  \includegraphics[width=\linewidth]{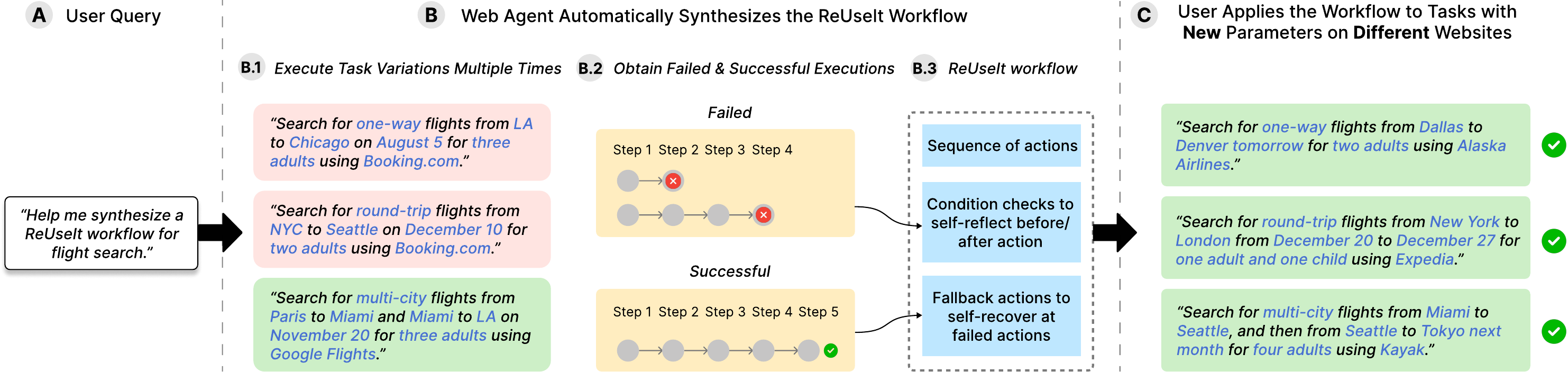}
  \caption{With \tool, users can reliably execute repetitive tasks using web agents guided by reusable workflows. The reusable workflow is synthesized based on the user query for a web automation task (A). The web agent then automatically generates tasks with various types of variations (B.1) and executes these variations multiple times to obtain both failed and successful runs (B.2). These attempts contribute to the reusable workflow that consists of a sequence of main actions, condition checks for self-reflection to identify issues, and fallback actions to self-recover at failed actions (B.3). Users can apply this workflow to guide the web agent to reliably execute new tasks on the same and different websites in the future (C).}
  \Description{This figure shows a flowchart for ReUseIt. Starting from the left, "A User Query" specifies a workflow synthesis query from the user. This query is passed to "B Web Agent Automatically Synthesizes the ReUseIt Workflow". In the workflow synthesis process, the web agent first "B.1 Execute Task Variations Multiple Times", then "B.2 Obtain Failed and Successful Executions", and finally obtains "B.3 ReUseIt Workflow". In the last step, the synthesized workflow is applied by the user to "C Tasks with New Parameters on Different Websites", for reliable web agent task execution.}
  \label{fig:teaser}
\end{teaserfigure}

\maketitle

\input{sections/1-introduction}

\input{sections/2-related_work}

\input{sections/3-formative_study}

\input{sections/4-system}

\input{sections/5-auto_evaluation}

\input{sections/6-user_evaluation}

\input{sections/7-discussion}

\input{sections/8-conclusion}

\begin{acks}
We would like to thank Jianfeng Gao, Michel Galley, Swadheen Shukla, and members of the Microsoft Research Deep Learning group for the brainstorming and discussion sessions that helped us shape our project. 
We want to thank our study participants for their participation and offering us valuable insights on the evaluation. 
We also want to thank the anonymous reviewers for their constructive feedback that helped us improve our paper. 
\end{acks}

\bibliographystyle{ACM-Reference-Format}
\bibliography{bibliography}

\appendix
\input{sections/appendix}

\end{document}

%% file: sections/0-abstract.tex
\begin{abstract}
AI-powered web agents have the potential to automate repetitive tasks, such as form filling, information retrieval, and scheduling, but they struggle to reliably execute these tasks without human intervention, requiring users to provide detailed guidance during every run. We address this limitation by automatically synthesizing reusable workflows from an agent's successful and failed attempts. These workflows incorporate execution guards that help agents detect and fix errors while keeping users informed of progress and issues. Our approach enables agents to successfully complete repetitive tasks of the same type with minimal user intervention, increasing the success rates from 24.2\% to 70.1\% across fifteen tasks. To evaluate this approach, we invited nine users and found that our agent helped them complete web tasks with a higher success rate and less guidance compared to two baseline methods, as well as allowed users to easily monitor agent behavior and understand its failures.
\end{abstract}

%% file: sections/1-introduction.tex
\section{Introduction}

\begin{figure*}[!ht]
    \centering
    \includegraphics[width=\linewidth]{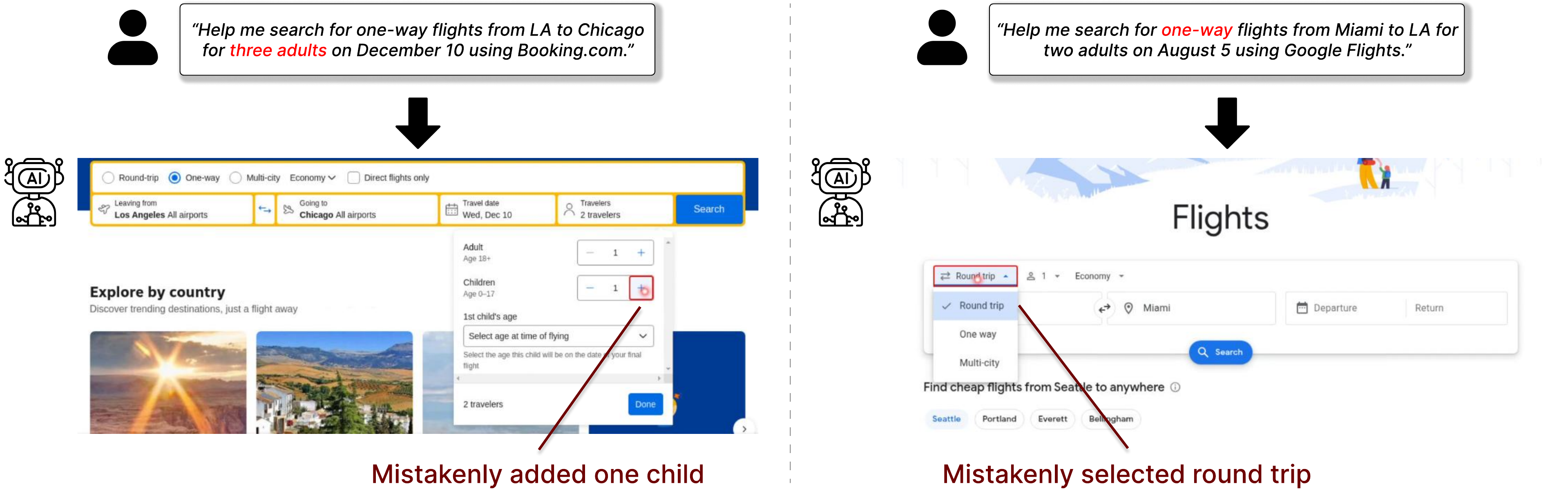}
    \caption{Given user instructions, web agents can make low-level but critical mistakes that prevent them from completing tasks. For example, the agent mistakenly used the ``+'' next to children to add adults on Booking.com (left), and mistakenly selected ``round trip'' from the drop-down menu despite the user asking for ``one-way'' flights on Google Flights (right).}
    \Description{This figure shows two examples of web agent mistakes. On the left, the web agent was asked to "search for one-way flights from LA to Chicago for three adults on December 10 using Booking.com." However, the agent mistakenly added one child, even though the task did not involve any child among the passengers. On the right, the web agent was asked to "search for one-way flights from Miami to LA for two adults on August 5 using Google Flights." However, the web agent mistakenly selected round-trip, which contradicts the required one-way flights.}
    \label{fig:introduction}
\end{figure*}

Data scientists, business analysts, and everyday users often need to perform repetitive interactions with the web that they may wish to automate. For example, a data scientist who wants to analyze housing markets needs to periodically scrape data from multiple online housing listings; a market researcher who hopes to understand changing customer attitudes towards certain products needs to collect information from multiple online forums; and, a customer who wishes to optimize their expenses to purchase a popular item needs to visit different websites to check availability and cost. AI-powered web agents have the potential to assist with these user needs that entail repetitive web tasks. Powered by vision language models (VLMs), web agents can perceive and take human-like actions on the web~\cite{openai2025operator, mozannar2025magentic}. With high-level inputs (prompts) from the user, web agents can autonomously interact with the web to navigate pages, submit forms, and collect information, even when they encounter dynamic web content (e.g., websites with changing layouts, pop-up menus) that traditional script-based automation techniques cannot handle~\cite{dong2022webrobot}.

However, despite their impressive capability, existing web agents struggle to reliably complete web tasks and apply a workflow for future reuse. From a high-level instruction, prior evaluation has shown that the state-of-the-art web agents achieved 66.0\% (Anthropic Sonnet 3.7 CUA) and 59.8\% (OpenAI CUA) accuracy to execute common web tasks~\cite{skyvern}. Even if the agent is provided with a successful workflow (e.g., guided by users during execution~\cite{mozannar2025magentic}), the agent struggles to reapply the workflow to the same task at a different time or to new tasks with similar workflows. In our benchmark evaluation of six common tasks, the average success rate of repeating the same tasks (five times each task) was 28.0\% and generalization to related tasks (five times each task) was 22.5\%, using the Magentic-UI agent framework~\cite{mozannar2025magentic}. The agent would sometimes hang, repetitively attempt wrong actions, or misunderstand the status of the workflow. Thus, to use agents for web automation, users currently need to interactively correct the agent's workflow every time they execute the web task, which presents a substantial limitation for the desired goal of automation. In fact, it can be difficult for users to watch agent actions on screen to effectively understand the progress of AI agents towards their given goals and provide feedback when they get stuck. Coupled with the fact (based on our formative user study, Section~\ref{sec:formative_user_study}) that users often struggle at pinpointing and offering guidance to resolve low-level but critical agent mistakes (e.g., cannot locate the right button to update the number of adult passengers, despite already understanding that it needs to update the passenger number, as illustrated in Figure~\ref{fig:introduction}), there are still substantial challenges to using AI agents for web task automation. 

To address these challenges, we propose an automatic workflow synthesis approach, \tool, that can produce reliably reusable workflows for AI agents. Motivated by prior research on learning and reusing workflows to improve agent execution performance~\cite{wang2024agent, li2024autoflow, yu2025survey}, our approach lets web agents automatically compose workflows from their successful and failed attempts with variational inputs that can be both (1) reliably used towards future repetitive tasks and (2) interpretable for users. Given a user input, such as Figure~\ref{fig:teaser}~A, instead of attempting to just complete the task, \tool first runs the task multiple times in parallel with different input variations that are generated based on the potential future reuse of the workflow, as shown in Figure~\ref{fig:teaser}~B.1 (in our experiments, the agent first repeats $n$=5 for the initial inputs and runs another $n$=5 for each different input variation). While the agent often cannot successfully complete all exploration runs, both successful and failed attempts provide valuable information that surfaces potential mistakes the agent can make during executions, especially common pitfalls, and recovery strategies (Figure~\ref{fig:teaser}~B.2). Based on these executions, \tool synthesizes an annotated workflow with \emph{execution guards} consisting of condition checks and fallback actions that describe the website state that the agent is expected to achieve before/after key interaction steps (identified based on prior failed execution traces) and what the agent should do if they find themselves in a different browser state (extracted according to successful experiences), as shown in Figure~\ref{fig:teaser}~B.3. With the synthesized workflow, users can apply it to reliably execute new tasks on different websites, as demonstrated in Figure~\ref{fig:teaser}~C. The execution guards also serve as the interface between the user and the agent, where users can examine the differences between the saved snapshot and the current workflow to track the progress and provide feedback. With this design, the burden on the user to track agent actions through lengthy action replays is turned into verification of key webpage states, providing users with a better experience in interacting with web agents.

To evaluate the effectiveness of \tool to generate reliable and interpretable workflows for web automation, we used Magentic-UI~\cite{mozannar2025magentic} as the agent framework and conducted (1) a benchmark evaluation on whether \tool's synthesized workflows can be reliably repeated on both the same and (unseen) task variations on fifteen tasks (three variations per task), and (2) a user study with nine participants to examine whether execution guards in the synthesized workflow can help users to understand and steer web agents. From the benchmark evaluation, web agents guided by the \tool's synthesized workflows achieved a 70.1\% success rate when executing repetitive tasks, improving the performance by 45.9\% over executing without the workflows. From the user study, the execution guards effectively helped users to identify and understand agent issues, and boosted their willingness and trust to use web agents in their own tasks. In sum, this work makes three main contributions: 
\begin{itemize}[leftmargin=*]
    \item A \textbf{formative study} that identifies the user challenges to detect and correct issues when interacting with web agents, and the need for more reliable and interpretable web agents. 
    \item An \textbf{automatic workflow synthesis approach}, \tool, that generates reliable and interpretable workflows for future reuse based on agents' past execution traces.
    \item A \textbf{benchmark evaluation} that demonstrates the effectiveness of \tool for repetitive web tasks and a \textbf{user study} showing how users can leverage the synthesized workflow to guide agents when solving challenging tasks. 
\end{itemize}

%% file: sections/2-related_work.tex
\section{Related Work} \label{sec:related_work}

Our work builds on top of prior research on LLM-based agents, web automation workflows, programming-by-demonstration techniques, and human-agent interaction.

\textbf{LLM-based web agents and their challenges.}
LLM-based web agents are autonomous systems that leverage LLMs to navigate websites and execute tasks based on natural language instructions. These agents translate the LLM's text outputs into web actions (e.g., click links and buttons, fill in forms, or web search) by specialized prompting and tool-use frameworks~\cite{openai2021webgpt, openai2025operator}. 
The usage of these agents has been explored in a wide range of scenarios, aimed at both professionals and end-users. 
On the professional side, web agents are being tested as digital ``office workers'' that could automate business workflows. For example, a recent benchmark, TheAgentCompany~\cite{xu2025theagentcompany}, simulated an enterprise intranet environment with internal web apps and data and evaluated LLM-based agents on everyday workplace tasks. 
On the consumer side, web agents can serve as advanced personal assistants that perform common web tasks on behalf of users. For instance, a web agent can book hotel rooms, search flight tickets, and reserve restaurants~\cite{yang2024agentoccam, zhou2023webarena}. More use cases include tracking product prices on e-commerce websites, retrieving answers from multiple forums, and summarizing daily feeds on social media~\cite{uzair2023autogpt, browsecomp2025}. 

Despite rapid progress, LLM-based web agents face significant challenges in repetitive tasks. 
A common concern is the reliability of task performance. Recent studies have found that only about 30\% of multi-step web tasks can be completed successfully by advanced agents like Operator~\cite{openai2025operator}, and many other agents achieve far lower success rates on average, often below 20\% on realistic web automation benchmarks~\cite{ning2025survey, xue2504illusion, skyvern, song2025bearcubs}. 
Another fundamental issue is inconsistent behavior that limits reusability. Since agent decisions emerge from probabilistic language model outputs, the exact sequence of actions can vary from run to run, and these agents fall short in being reused in similar tasks~\cite{xue2504illusion, openai2025operator, pan2024webcanvas, levy2024st}. 
Such ad-hoc behavior also leads to unpredictable failures. When an unexpected webpage status change happens (e.g., a pop-up menu), the agent behaves differently across runs and fails in various ways~\cite{ledel2025fundamental}. 
To address these limitations, we focus on augmenting agent reliability and consistency with execution guards and self-recovery, and boosting reusability through successful and failed experiences from multiple executions of the same and similar tasks. 

\textbf{Workflow synthesis for web agents.}
A promising approach to address web agents' limitations is to automatically learn and reuse execution workflows as part of their reasoning process. For example, Agent Workflow Memory (AWM)~\cite{wang2024agent} induces commonly used workflows from an agent's past experiences and injects them into the agent's memory to guide future actions. Leveraging learned workflows, AWM significantly boosted success rates on long-horizon web tasks (e.g., +24.6\% on the Mind2Web benchmark~\cite{deng2023mind2web}) while reducing the number of execution steps. Other work has explored representing workflows in various forms to guide the agent, including natural language~\cite{li2024autoflow}, descriptive workflow language~\cite{deng2023mind2web}, and pseudo-code~\cite{zhou2023webarena, wang2024agent}. 

Nevertheless, the reusability of execution workflows has been less explored by prior research. Most of these studies analyze the task success rate of agents equipped with synthesized workflows, but offer limited insights into agent performance when reusing these workflows in repetitive and similar tasks. In our work, we induce agent workflows automatically, with a specific focus on synthesizing execution guards in the workflow to enhance agent reliability and reusability using any representation of the workflow. 

\textbf{Programming by demonstration for web automation.}
Programming by demonstration (PbD) techniques have also been applied to address web automation challenges via scripting. In this paradigm, a human user performs a web task step-by-step, and the system synthesizes a deterministic program that replicates those actions. For example, WebRobot~\cite{dong2022webrobot} constructs web automation scripts by observing user interactions with a web browser. DiLogics~\cite{pu2023dilogics} leverages natural language processing to generalize user demonstrations into web automation scripts that handle diverse input conditions. MIWA~\cite{chen2023miwa} introduces a mixed-initiative interface for web automation, providing visual explanations and interactive debugging tools to help users understand and refine the generated scripts. In particular, programming by navigation synthesis~\cite{lubin2025programming} has been proposed to support the iterative user specification refinement process. 

However, a major limitation of PbD approaches is the heavy human effort required, as the user needs to provide careful demonstrations and sometimes corrections for each new task~\cite{sosa2022learning}. Furthermore, the deterministic nature of these synthesized programs means that they cannot easily be generalized to variations of tasks in the future, despite the overall workflow being similar. In contrast, our approach avoids manual demonstrations. Instead, we let the web agent learn through its own trial-and-error to automatically synthesize execution guards from its failures and successes that can generalize beyond the initial task. 

\textbf{User interaction with AI agents.}
To enable users to inspect, interrupt, and redirect AI agents, researchers have explored interaction design that keeps humans in the loop during agent execution. Recent systems, such as Magentic-UI~\cite{mozannar2025magentic}, Cocoa~\cite{feng2024cocoa}, and Plan-Then-Execute~\cite{he2025plan}, exemplify human interaction with web agents. They allow users to co-plan with web agents, observe real-time agent actions, take over tasks when agents get stuck, and approve/reject critical agent executions.
In addition to web agents, recent research has improved the performance of and studied human interaction with agents in other types of graphical user interfaces (GUIs), such as mobile UIs~\cite{zhang2025appagent, wang2024mobile, hao2025uncertainty, yang2025ferret}, specific applications~\cite{li2023sheetcopilot, liang2025tabletalk}, computer operating systems~\cite{wu2024copilot, xie2024osworld}, and games~\cite{wang2023voyager}.
By investigating the impact of human involvement during AI agent execution, prior work has found improved user understanding, preference alignment, and agent safety. For instance, Uncertainty-Aware GUI Agent~\cite{hao2025uncertainty} can detect agent uncertainty in decision phases and request user feedback in ambiguous states; the Dark Patterns study~\cite{tang2025dark} shows that human oversight can help reduce errors when agents face deceptive interfaces; the Human-Centered Evaluation framework~\cite{chen2025toward} suggests incorporating human evaluators into GUI agent assessment pipelines to ensure trustworthiness and alignment. 

Following this idea, our work loops users into the agent's execution process, such that users can identify and understand agent issues as they occur. Specifically, when the agent is uncertain about how to proceed or encounters a failure that it cannot recover from, it invites user guidance with issue analysis, aiming for more interpretable, trustworthy, and controllable agents.

%% file: sections/3-formative_study.tex
\section{Motivation}
To understand the gap between current web agents and user needs for automating repetitive web tasks, we first evaluated the performance of web agents on a set of six common web automation tasks, and conducted a formative study to examine user experience of using these web agents.

\subsection{Preliminary Evaluation: Reliability of web agents to execute common web automation tasks} \label{sec:prelim_eval}

\begin{figure*}[!ht]
    \centering
    \includegraphics[width=.8\linewidth]{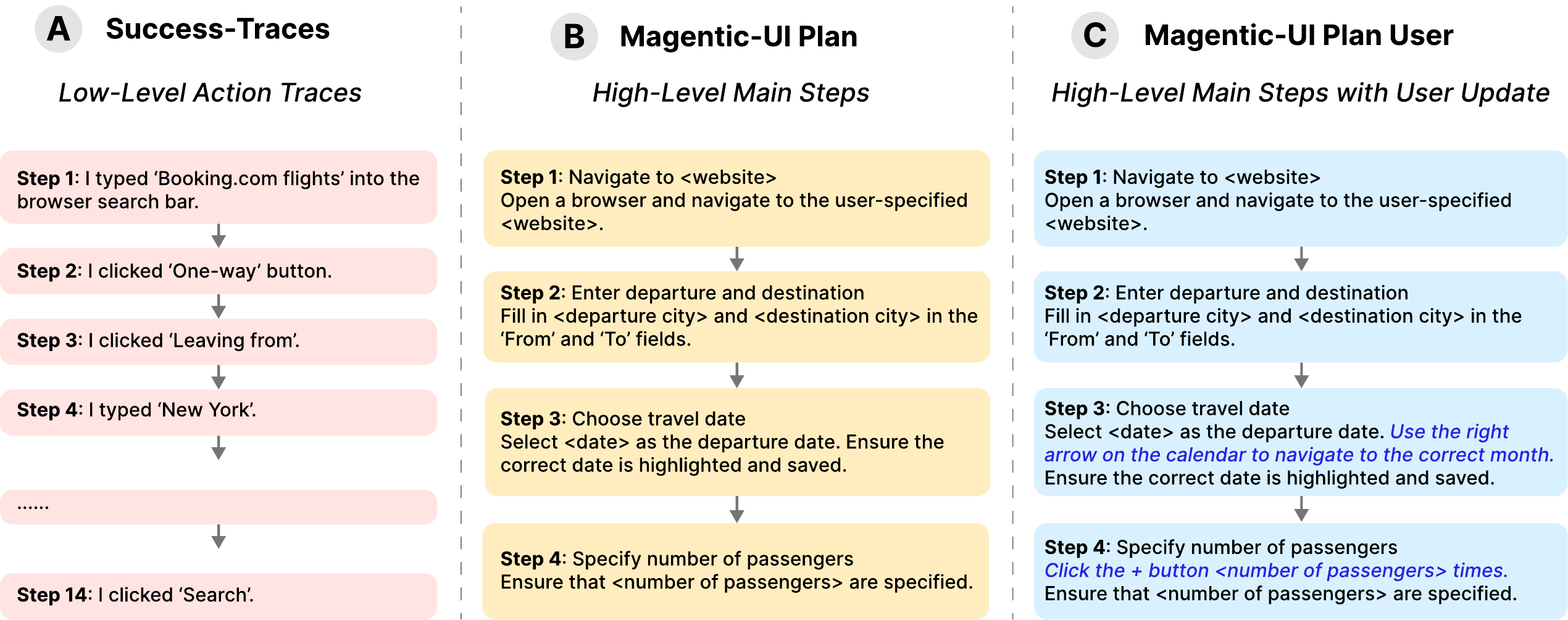}
    \caption{Examples of Success-Traces (A), Magentic-UI Plan (B), and Magentic-UI Plan User (C; user edits are shown in blue text).}
    \Description{This figure shows three examples of action traces/workflows. On the left, it shows "A Success-Traces", where a list of detailed agent actions is outlined, such as text typed and buttons clicked. In the middle, it displays "B Magentic-UI Plan", where a list of high-level steps containing the main actions to take is shown, such as navigating to a website and entering user-specified information. On the right, it illustrates "C Magentic-UI Plan User", which is largely the same as "B Magentic-UI Plan" but with user edits, such as action cues to help the agent recover from failures.}
    \label{fig:workflow_baselines}
\end{figure*}

To quantify overall agent performance on common web automation tasks, we collected six tasks from prior work~\cite{he2024webvoyager, deng2025simura,krosnick2021understanding, dong2022webrobot}. These tasks require web agents to navigate the web, fill forms, apply filters, extract information, and summarize content, and the tasks span different application domains, including searching flights, housing, pets, online shopping products, research publications, and news. More details of the tasks are in Appendix~\ref{appendix:formative_tasks}. We evaluated the web agent's performance to complete these tasks under three conditions, to understand the agent's ability to finish the tasks and reapply previous low-level action traces and high-level plans towards new tasks:

\begin{itemize}[leftmargin=*]
\item \textbf{Condition 1 Task-Only:} The agent is only provided with the task prompt from the user (e.g., ``help me search for product information from an online shopping website''); 
\item \textbf{Condition 2 Task + Success-Traces:} The agent is provided with the user prompt describing the task together with a previous successful execution trace containing low-level, detailed actions taken on each webpage (e.g., opened a drop-down menu, typed content in a text field; Figure~\ref{fig:workflow_baselines}~A); 
\item \textbf{Condition 3 Task + Magentic-UI Plan:} The agent is provided with the task prompt together with a high-level workflow outlining the main steps obtained with Magentic-UI's plan-learning method (e.g., step 1: navigate to a user-specified website, step 2: search content; Figure~\ref{fig:workflow_baselines}~B). 
\end{itemize}

To measure the agent's performance, we used the success rate ($SR$) as the evaluation metric. $SR$ is defined as the number of successful executions out of the total number of runs. For each task, we ran five executions and computed the fraction of successful runs. We also generated variation tasks (with variations on task attribute, category, and website; more details are presented in Section~\ref{sec:auto_workflow_synthesis}) and executed each variation five times to test workflow reuse on closely related tasks. In total, we executed each task and its variations 20 times, and report the mean $\pm$ std $SR$ (in $\%$) across all these executions. To determine whether the executions were successful, we manually inspected the screen recordings to judge if the agent had satisfied all user requirements and found the intended results. 

\begin{figure*}[!ht]
    \centering
    \includegraphics[width=.9\linewidth]{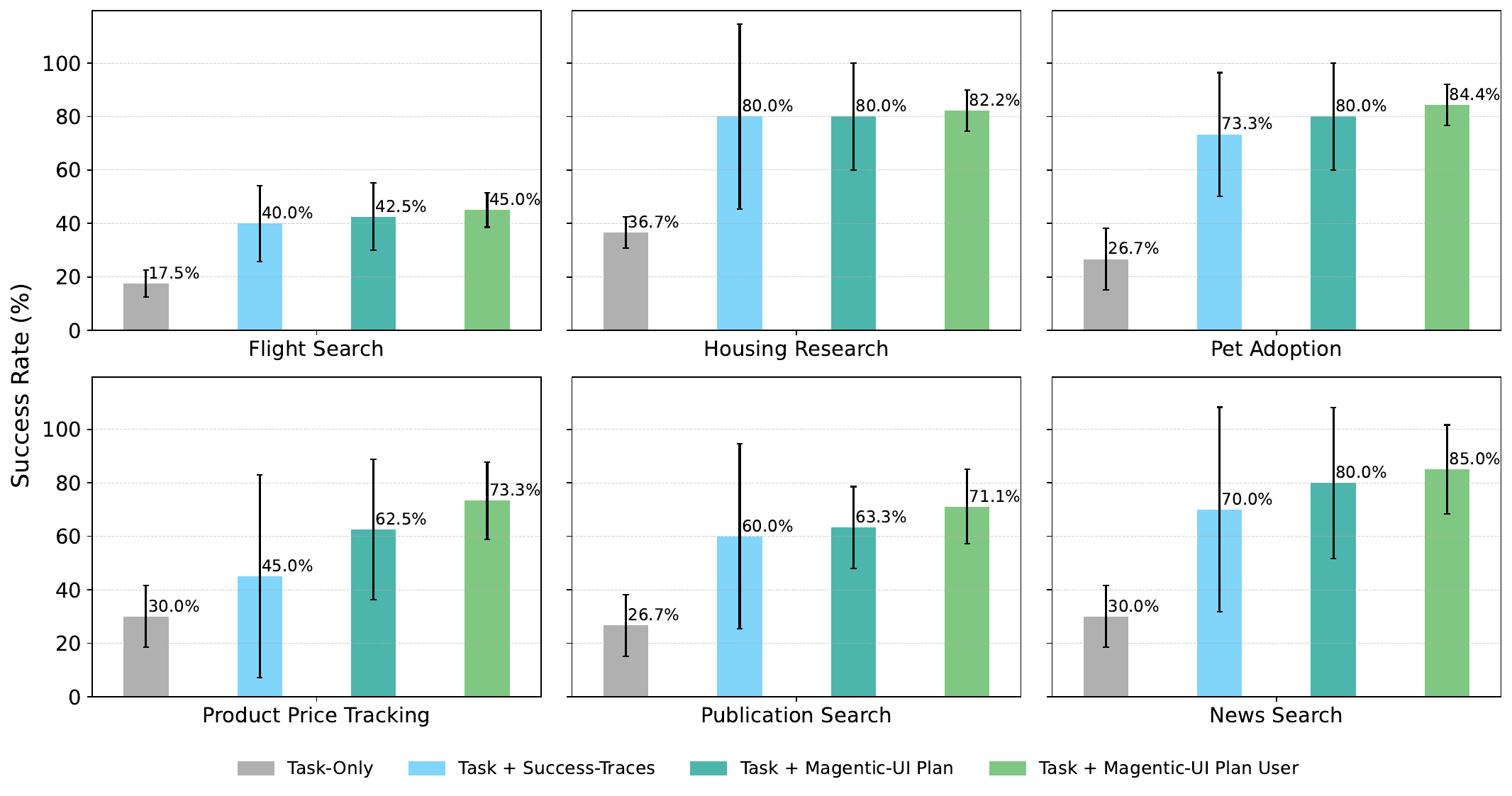}
    \caption{Evaluation of web agents' success rate on six web automation tasks and their variations (avg $\pm$ std success rate).}
    \Description{This figure shows six bar charts with task success rates for the web automation tasks.}
    \label{fig:user_eval_grid}
\end{figure*}

\paragraph{Evaluation Results}
Figure~\ref{fig:user_eval_grid} summarizes the experiment results. Across tasks, the $SR$s are: Task-Only ($27.9\% \pm 6.3\%$), Task + Success-Traces ($61.4\% \pm 16.1\%$), and Task + Magentic-UI Plan ($68.1\% \pm 15.1\%$). In addition, we report the mean $SR$s with 95\% confidence intervals of each task and across all tasks in Appendix~\ref{sec:prelim_eval_sr_mean_ci}. These results show that (1) although each task succeeded at least once, the agent's performance is inconsistent across runs and the average $SR$ remains low; and (2) while successful execution traces and plans can greatly improve the $SR$ for repeating the same or similar tasks, a gap remains before users can fully hand off web interaction tasks to agents. 

\subsection{Formative Study: Challenges for users to apply web agents in web automation tasks} \label{sec:formative_user_study}
Given that web agents do not guarantee high success rates on web tasks, we conducted a formative study to examine: 

\begin{itemize}[leftmargin=*]
\item Can users identify the issues of web agents, how do they find issues, and what makes diagnosis challenging?
\item How do users provide guidance to the agent when it fails, which aspects of this process do they find hard, and can the guidance effectively help the agent execute repetitive tasks?
\end{itemize}

\paragraph{Study Design}
We recruited six participants from a large company (1 female, 5 male; mean age = 26.33, SD = 1.11; IDs S1-P1 to S1-P6). From a pre-study questionnaire, all participants used LLMs daily, and their self-reported prompting experience showed that three of them were experts at writing advanced prompts to address complex tasks, two were good at intermediate prompting techniques, and one was familiar with basic prompting for simple daily tasks (Table~\ref{tab:formative_participants_demographics}). 

\begin{table*}[!ht]
    \centering
    \footnotesize
    \caption{Participants' self-reported frequency of LLM use, expertise in basic, intermediate, and advanced prompting (1=strongly disagree, 5=strongly agree), and time taken to find agent issues and provide guidance.}
    \begin{tabular}{ccC{1.5cm}C{1.5cm}C{1.5cm}cc}
        \toprule
        \multirow{2.5}{*}{\textbf{ID}} & \multirow{2.5}{*}{\textbf{Frequency of LLM Use}} & \multicolumn{3}{c}{\textbf{Prompting Expertise}} & \multirow{2.5}{*}{\textbf{Find Issues}} & \multirow{2.5}{*}{\textbf{Provide Guidance}} \\
        \cmidrule(){3-5}
         &  & \textbf{Basic} & \textbf{Intermediate} & \textbf{Advanced} \\
        \midrule
        S1-P1 & Daily & \Likert{5} & \Likert{5} & \Likert{5} & $923$ s & $988$ s\\
        S1-P2 & Daily & \Likert{5} & \Likert{4} & \Likert{4} & $952$ s & $997$ s\\
        S1-P3 & Daily & \Likert{5} & \Likert{5} & \Likert{4} & $898$ s & $858$ s\\
        S1-P4 & Daily & \Likert{5} & \Likert{5} & \Likert{5} & $867$ s & $902$ s\\
        S1-P5 & Daily & \Likert{4} & \Likert{4} & \Likert{2} & $973$ s & $933$ s\\
        S1-P6 & Daily & \Likert{5} & \Likert{5} & \Likert{5} & $885$ s & $932$ s\\
        \bottomrule
    \end{tabular}
    \Description{This table summarizes the demographic information of the formative study participants, as well as the time they took to find agent issues and offer guidance to help the agent recover from issues during the study.}
    \label{tab:formative_participants_demographics}
\end{table*}

During the study, after a brief introduction and example task walkthrough ($\sim$10 mins), each participant was asked to inspect three web automation tasks that web agents failed to complete (we shuffled the task order and ensured that each task was examined by three different participants; $\sim$30 mins) and offer guidance to the agent to help it recover. In each task, participants were provided with slides containing the task description, the screen recording of agent actions, and the workflow the agent followed in the form of a list text block. The workflow was generated with Magentic-UI's approach~\cite{mozannar2025magentic}. 
Participants were invited to verbally analyze why and how the agent failed to execute the tasks and edit the workflows on the slides. Finally, we conducted a semi-structured interview with each participant to get a deeper understanding of their user experience ($\sim$20 mins). More details of the formative study materials can be found in Appendix~\ref{appendix:formative_interview}.

We performed thematic analysis~\cite{braun2006using, braun2012thematic} using the qualitative data from interviews to code participants' experience in spotting and fixing agent errors. We then performed a follow-up experiment to evaluate the web agent's performance in completing the tasks guided by these user-fixed workflows. 

\paragraph{Study Result 1: User Experiences} \label{sec:formative_results_user}
In general, we noticed that users spent significant time (around half of the task time) finding agent issues and were not fully certain about some of the causes. In addition, users took great efforts to address low-level but critical agent mistakes when offering guidance. 

\textbf{User detection of agent issues.}
From the study, we found that the participants used different ways to identify agent issues. All participants were able to find agent issues by examining the screen recordings. The time taken by them to find issues is reported in Table~\ref{tab:formative_participants_demographics}. They frequently paused and replayed the recordings to check agent actions, such as how the agent handled pop-ups, applied filters, filled forms, and navigated pages. They described that the recordings helped them pinpoint problems (S1-P3) and understand the agent's intent (S1-P6).
Additionally, participants also cross-referenced the agent's workflows (S1-P2: \textit{``I referenced the workflow to see the step breakdown and then came back to the video to check the actual actions.''}), and used the workflows to infer potential causes of issues, such as missing instructions on how to apply filters (S1-P4: \textit{``It [the agent] might not know the possible actions it can take... so it's relying very much on keyword-based search.''}).

The agent issues found by our participants were mostly low-level but critical mistakes, such as: (1) Filters not applied. The agent proceeded without applying all the requested filters. (2) Navigation challenges. The agent was stuck by pop-up banners, new pop-up UI elements like drop-down menus triggered by clicking a text field. (3) Random clicks/scroll. The agent clicked on something random or did not have a good awareness of the page layout, scroll position, or visible elements. (4) Ambiguity about intent. The agent was unclear about certain user instructions, such as ``first/top/most recent''. (5) Partial actions. The agent failed to finish clicking the search button or set all required parameters. 

While some errors were relatively easy to diagnose when symptoms were visible (e.g., a missing filter or an incomplete action), other cases were more ambiguous. Diagnosing became harder when tasks involved subjective goals, when site-specific glitches obscured the cause of agent failure, or when the agent took unexpected actions (S1-P2: \textit{``I don't think this agent actually did anything related to this filter... it just picked one of the random ones that I never specified.''}, S1-P5: \textit{``maybe the search box where it [the agent] seems to have clicked and nothing happened... it could be a website glitch, but hard to say.''}, and S1-P6: \textit{``I don't know how it comes to this selection. I didn't see it [an article] even on the web page.''}).

\textbf{User guidance to the agent.} \label{sec:formative_results_user_strategies}
When the agent made mistakes, we observed that the participants used a combination of different strategies to provide guidance, which can be summarized into five categories: 
(1) Improve the specificity of instructions. A common strategy was to add details to the workflow and specify each expected agent action (S1-P1, S1-P2, S1-P5). 
(2) Decompose instructions into clear steps. To ensure reliable execution, participants translated complex instructions into explicit and clear steps for the agent to follow (S1-P2, S1-P6). 
(3) Add action verification for self-reflection. Participants highlighted building guardrails that confirm each critical action before proceeding, and branching to take alternative actions when something goes wrong so mistakes do not silently cascade (S1-P1, S1-P3). 
(4) Provide action candidates for agent reference. Participants focused on reducing ambiguity by surfacing explicit action options the agent can choose from, such as available UI operations and their effects (S1-P4, S1-P6). 
(5) Include tips to handle webpage changes or common issues. To keep agent workflows resilient against varying webpage changes or recurring agent issues, participants embedded generic troubleshooting cues in the workflow (S1-P1, S1-P4). 

Most participants focused on adding instruction details and clarifying required steps to reduce ambiguity. However, they expressed uncertainty regarding what level of specificity in the instruction was good for the agent to effectively process and follow (S1-P1, S1-P5). In addition, they mentioned that improving the specificity and clarity of instructions can be time-consuming, so they may perform the tasks manually instead (S1-P2: \textit{``I'd hope the AI could generate a prompt like this, and then it can do the task. Cuz I don't really want to spend a lot of time on writing up this prompt since I would have already been able to do that myself.''}). The time taken for participants to provide guidance is summarized in Table~\ref{tab:formative_participants_demographics}.

\paragraph{Study Result 2: Effectiveness of User-Fixed Workflows} 
To understand the effectiveness of user-provided guidance (e.g., Figure~\ref{fig:workflow_baselines}~C), we conducted a follow-up experiment with the user-fixed workflows, and the results are shown in Figure~\ref{fig:user_eval_grid} (\textbf{Task + Magentic-UI Plan User}). The success rate in this figure was averaged across users, and Appendix~\ref{appendix:formative_followup_results_full} reports the full results for each user. Across tasks and user-fixed workflows, Magentic-UI Plan User has improved the $SR$ by around $5.4\%$ compared to Magentic-UI Plan. From these results, the strategies that most aligned with higher and more consistent success were decomposing instructions into clear steps, and providing action candidates and recovery tips (S1-P4 and S1-P6 using these strategies achieved the top overall $SR$). A mix of adding specificity, action verification, and recovery tips by S1-P1 also yielded comparatively high $SR$s. 

\subsection{Reflection: Gaps in applying web agents for repetitive web tasks}
Our preliminary evaluation and formative study results show that while users can interactively work with web agents to complete tasks, their efforts spent on helping web agents might outweigh the benefits from agent automation. Specifically, we noticed the following three gaps for users to apply web agents in repetitive tasks.

\textbf{Substantial time needed to understand issues.}
All our participants devoted considerable time to reviewing screen recordings and monitoring the agent's step-by-step actions to locate failures and reason about their causes. When actions were ambiguous, participants expressed uncertainty and frequently sought confirmation of their interpretations (S1-P1, S1-P2, S1-P3, S1-P5). This indicates that even careful inspection sometimes left them unsure whether their understanding of the issues was correct.

\textbf{Significant effort to fix low-level but critical mistakes.}
Examining the edits made to the workflows, we observed that participants consistently targeted the agent's low-level but critical action issues. Despite these efforts, user guidance improved the success rate by only a small margin according to our follow-up experiments. Participants also questioned whether they would want to invest this level of effort in their own tasks (S1-P2, S1-P5).

\textbf{Users wanted their guidance to be effectively reused.}
Participants preferred web agents that they could reuse without having to re-specify guidance each time. For example, they pointed to tasks like flight booking (S1-P2), course registration (S1-P2), job search (S1-P3), social media/news/paper summarization (S1-P1, S1-P5, S1-P6), housing search (S1-P3, S1-P6), and product price search (S1-P4, S1-P5). In these use cases, they were willing to invest in workflow improvements so long as those edits would carry forward to future runs and indeed help to improve the agent reliability.

%% file: sections/4-system.tex
\section{System Design} \label{sec:system_design}
This section introduces the system we designed and the technical details of our implementation. Our system is designed with the following design goals (DGs) to address the gaps we learned from our formative study.

\begin{itemize}[leftmargin=*]
\item \textbf{DG1: Automatically mitigating low-level but critical agent issues.}
Low-level but critical agent issues (as introduced in Section~\ref{sec:formative_results_user}) are often out of users' expectations and can be hard to spot. Instead of relying on users to fix these errors hands-on, the web agent should avoid these errors and recover autonomously at runtime. Our approach is to let web agents automatically synthesize workflows from their successful and failed experiences to avoid repeating mistakes. We further develop an agent self-reflection mechanism for the agent to analyze web states to detect and recover from issues during execution (Section~\ref{sec:auto_workflow_synthesis}). 

\item \textbf{DG2: Surfacing issues to help users identify where and why agents fail.}
In cases where agents fail to find or fix minor issues, the system should provide users with additional information explaining \emph{where} and \emph{why} the failure occurs. Thus, users can act accordingly to take control and fix issues without replaying agent actions frame-by-frame to speculate the causes. As users provide feedback to help agents recover from issues, their feedback should be incorporated as part of the workflow to prevent error recurrence. We continuously update the workflows with user guidance and propagate them to future task executions (Section~\ref{sec:workflow_guided_execution}). 
\end{itemize}

\subsection{Automatic Workflow Synthesis (DG1)} \label{sec:auto_workflow_synthesis}
To allow the web agent to locate and correct issues on its own, we propose an automatic workflow synthesis approach to guide task execution. Inspired by the user strategies from the formative study, we synthesize workflows that provide detailed steps, webpage state checks, and recovery tips for the agent to follow specific instructions, self-reflect to detect issues, and self-recover when issues occur. 

\begin{figure*}[!ht]
    \centering
    \includegraphics[width=\linewidth]{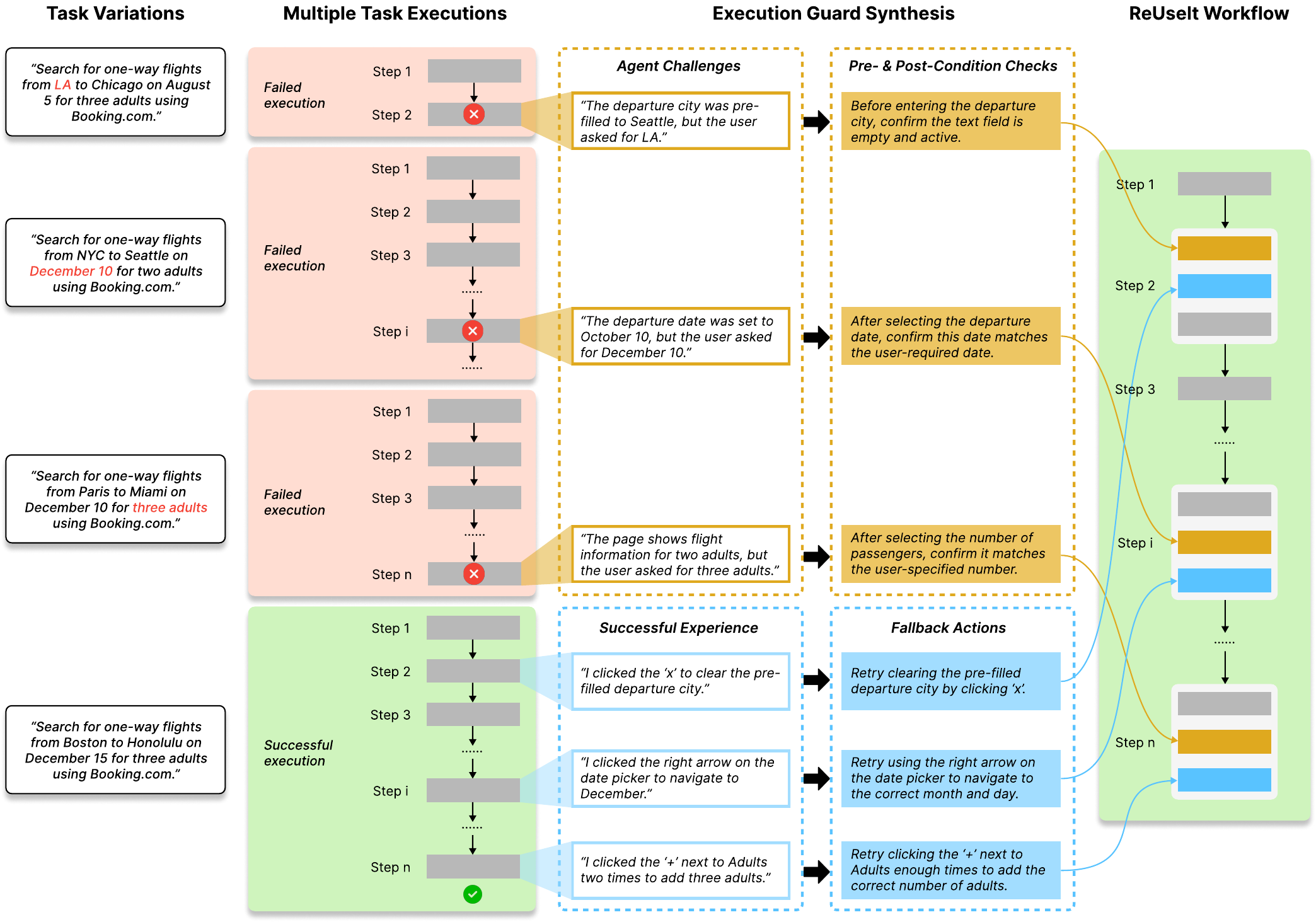}
    \caption{Automatic workflow synthesis pipeline. From multiple task executions of tasks with variations, \tool extracts agent challenges and successful experiences from failed and successful executions, respectively. Agent challenges are turned into condition checks at failed steps, while successful experiences contribute to fallback actions for agent self-recovery. Condition checks and fallback actions are concatenated with the corresponding step to form a unit in the synthesized workflow. Note that specific values are turned into generic framing in the workflow to support task generalizability.}
    \Description{This figure shows a flowchart for the automatic workflow synthesis pipeline. Starting from the left, under "Task Variations", there are four task variations. These variations are executed multiple times to obtain failed and successful attempts under "Multiple Task Executions". These attempts are used to synthesize execution guards, as shown under "Execution Guard Synthesis". Specifically, failed attempts reveal challenges faced by the agent and are turned into condition checks; successful attempts identify successful experience and are converted into fallback actions. Lastly, these synthesized execution guards are concatenated with the workflow structure to form the synthesized ReUseIt workflow.}
    \label{fig:automatic_workflow_synthesis}
\end{figure*}

In our approach, we synthesize workflows from both failed and successful agent executions. Failed executions can help expose common agent challenges that are then converted into \textbf{pre- and post-condition checks} around each action. For example, before navigating to a new page, the agent should verify that all required form fields are complete and correct; after entering a user-specified value in a text field, the field should display exactly that value.
Successful executions, in turn, yield \textbf{fallback actions}, the recovery strategies for steps where failures commonly occur, and \textbf{workflow structure}, the sequence of major steps to which the derived condition checks and fallback actions are attached. The action details of a step and the corresponding condition checks and fallback actions form a unit, which is concatenated with the rest of the steps/units to obtain the synthesized \tool Workflow. Figure~\ref{fig:automatic_workflow_synthesis} illustrates this workflow synthesis process.

\textbf{Multiple task executions.}
To synthesize the workflow, the first step is letting the web agent execute web tasks multiple times. Here, the tasks include the original task (directly from the user's task description), as well as the variation tasks generated from the original task. The goal of adding variation tasks is to enable the synthesized workflow to generalize across closely related tasks. We adopt a systematic way to generate these variation tasks based on \texttt{attribute}, \texttt{category}, and \texttt{website} variations. These variations exercise different levels of agent generalizability: \texttt{attribute} variation requires value substitution on the same webpage; \texttt{category} variation involves tab/category switches within a webpage; \texttt{website} variation includes UI layout change on different websites. For example, for the flight search task \textit{``search the <flight type> <cabin type> ticket from <departure city> to <destination city> on <date> for <number of passengers> on <website>''}, \texttt{attribute} variations include <cabin type> (e.g., economy, business), <departure city>, <destination city>, <date>, and <number of passengers>; \texttt{category} variations include <flight type> (e.g., one-way, round-trip); \texttt{website} variations varies different flight search <website> (e.g., Booking.com, Expedia.com). 
To generate the variation tasks, \tool uses the \textbf{variation task generation prompt} (Appendix~\ref{sec:prompt_var_task_gen}) to guide the LLM to automatically instantiate task variants. With the original task, the LLM identifies the values that can be substituted and groups these values into the \texttt{attribute}, \texttt{category}, and \texttt{website} variation categories. Within each category, the LLM needs to generate one variation task. We set the number of task executions as five for both the original and variation tasks, as this can lead to at least one successful execution for workflow synthesis based on our experiments. 

\textbf{Execution guard synthesis.} \label{sec:execution_guard_synthesis}
Multiple task executions yield failed and successful executions. For the failed attempts, we obtain the agent messages, which describe each agent action (e.g., navigated to a webpage, clicked a button, entered a value). 
For successful attempts, we likewise capture the agent's messages and additionally use Magentic-UI's \textit{plan learning} module~\cite{magui_planlearning} to synthesize a high-level plan consisting of the major steps. This plan serves as the workflow structure of the \tool Workflow. The prompt for plan learning was borrowed from Magentic-UI~\footnote{\url{https://github.com/microsoft/magentic-ui/blob/main/src/magentic_ui/learning/learner.py}}. 
Leveraging the messages and plans from multiple task executions, we synthesize the execution guards---condition checks and fallback actions---to augment the workflow structure. First, \tool uses the \textbf{condition check synthesis prompt} (Appendix~\ref{sec:prompt_condition_check}) to guide the LLM to extract agent challenges from failed execution messages and convert these challenges into conditions that should be met. From error messages, such as \textit{``failed to, didn't, or couldn't''} perform a certain action because \textit{``a button is inactive, a text field can't be located, or a page doesn't load,''} the LLM extracts the error pattern and convert it into a pre- or post-condition check for this action in the form of \textit{``Before/After doing <action>, ensure <condition> is met.''}
Next, \tool assembles the \textbf{fallback action synthesis prompt} (Appendix~\ref{sec:prompt_fallback_action}) to ask the LLM to find agent actions from successful execution messages at failed steps and turn these successful experiences into recovery strategies. Based on the description of agent actions, such as \textit{``I clicked, I navigated to, or I typed <UI element>,''} the LLM turns these actions into retry strategies, such as \textit{``Retry performing <action> by clicking, navigating to, or typing <UI element>.''}
Finally, \tool assembles the \textbf{workflow synthesis prompt} (Appendix~\ref{sec:prompt_workflow_syn}) to insert the synthesized condition checks and fallback actions into the corresponding step in the workflow structure to form the \tool Workflow. 

\textbf{Remarks.} 
Our current approach allows a parallel workflow synthesis from both successful and failed experiences simultaneously. This method can be altered into a sequential alternative if the successful execution can be quickly obtained, e.g., within 3 runs. Based on this successful execution, we can obtain the workflow structure and iterate to add condition checks and fallback actions with further executions. For tasks that are challenging for the agent to easily obtain a successful execution, the parallel approach is more efficient since users do not need to wait long for a successful run. We used the parallel approach in our implementation to accommodate tasks with multiple difficulty levels. 

\subsection{Agent Execution Guided by Synthesized Workflow (DG2)} \label{sec:workflow_guided_execution}

\begin{figure*}[!ht]
    \centering
    \includegraphics[width=\linewidth]{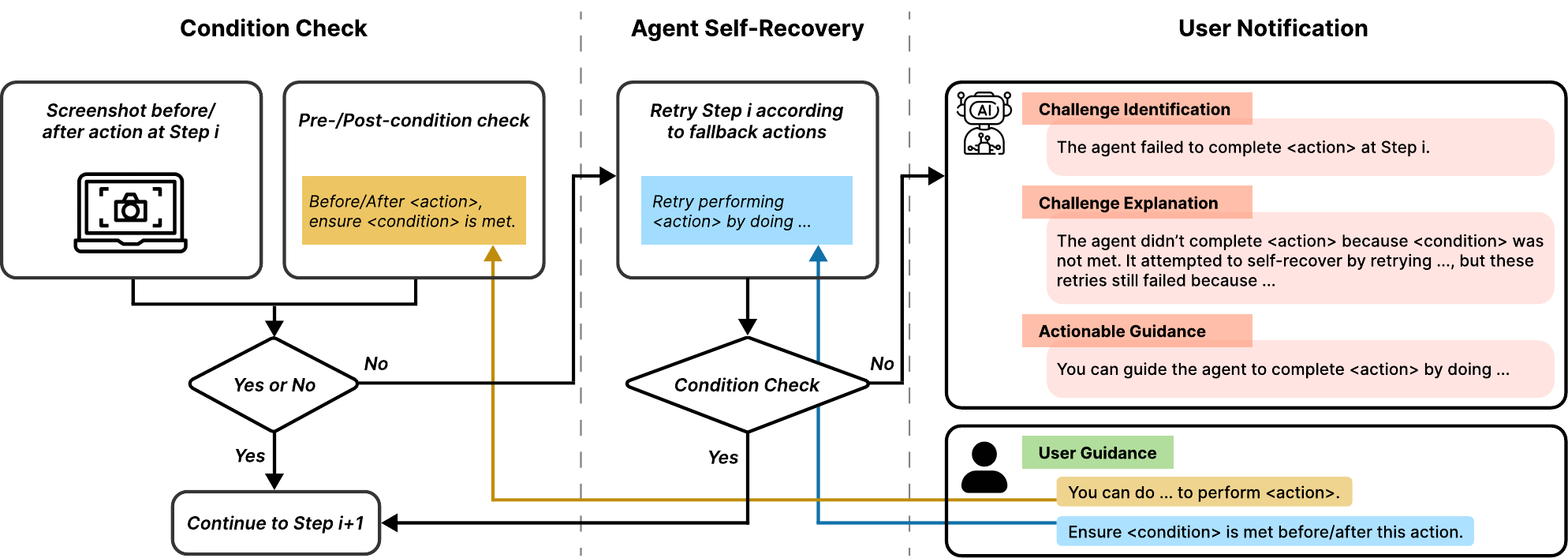}
    \caption{Agent execution guided by the synthesized workflow. At actions where pre-/post-condition checks exist in the workflow, \tool captures the webpage screenshot and does a Q\&A to determine if the condition check is met before/after this action. If the answer is ``Yes'', the agent proceeds to the next action, but if not, it retries executing this action according to the fallback actions. The agent moves on to the next action if the condition check is met after retries. However, if the retries fail, \tool notifies the user by providing a challenge analysis to help them understand the issue and provide guidance. User guidance is used to iteratively update the condition checks and fallback actions in the workflow.}
    \Description{This figure shows a flowchart for agent execution guided by the synthesized workflow. In the left column, ReUseIt performs "Condition Check" before/after step i by comparing the webpage screenshot and the synthesized condition checks to produce a Yes/No answer. If Yes, the agent moves on to step i+1; if No, the agent is guided to perform "Agent Self-Recovery", which is shown in the middle column. To self-recover, the agent retries step i according to the synthesized fallback actions and does the condition check again after the retries. If the condition is met, the agent continues to step i+1, but if not, the agent needs to send a "User Notification" message, as shown in the right column. In this message, the agent identifies and explains its challenge, as well as offers the user actionable guidance that can help the user offer feedback to unblock the challenge. The user feedback is used to iteratively update the synthesized condition checks and fallback actions.}
    \label{fig:condition_check_user_alert}
\end{figure*}

The execution guards in the synthesized workflow guide agent execution by asking the agent to reflect before and after each action and self-recover if it fails at a certain action. If the self-recovery fails, the agent notifies the user with issue analysis to obtain guidance. Figure~\ref{fig:condition_check_user_alert} shows this pipeline. 

\textbf{Condition check and agent self-recovery according to fallback actions.}
With synthesized workflows, \tool performs condition checks before/after each agent action. For each check, it uses the \textbf{condition check prompt} to evaluate if the expected conditions are met. First, \tool captures the webpage screenshot before/after each action. It then pairs the screenshot with the condition check, and asks the LLM to produce a ``Yes/No'' answer along with an explanation. If the condition is met, the agent moves on to the next action; if not, it self-recovers by re-executing the failed step up to three times according to the fallback actions. 

\textbf{User notification.}
If the agent's self-recovery fails, \tool iden\-tifies and explains the issue to notify the user. It uses the \textbf{challenge identification and explanation prompt} to instruct the LLM to analyze the condition check, as well as agent messages and screenshots from all attempts, in order to produce three aspects of content: (1) \emph{where} the failure occurred; (2) \emph{why} the agent failed; (3) \emph{what} agent behaviors caused the failure. The LLM first refers to the condition check to locate at which step the agent failed. Next, it explains why the agent's actions were considered failed based on the conditions that were not met. In addition, it summarizes agent messages and webpage screenshots and analyzes what specific actions led to the failure. 
Lastly, \tool offers actionable tips to help the user provide guidance to unblock the agent's progress. It assembles the \textbf{actionable guidance prompt} to guide the LLM to summarize the fallback actions at failed steps and outline suggestions on \emph{how} to help the agent recover for the user's reference.

\textbf{User guidance that iteratively updates execution guards.}
With user-provided guidance, \tool updates the execution guards through the \textbf{user guidance integration prompt}. This prompt asks the LLM to parse user guidance into condition checks or fallback actions. For instance, the content associated with \textit{``You need to make sure <condition> is met before/after <action>''} should be added to condition checks; while specific actions, such as \textit{``You can click <UI element> to perform <action>''} should be integrated into fallback actions. 
By iteratively incorporating user guidance, the automatically synthesized workflow ``learns'' with human input over time to strengthen the execution guards. 

\subsection{Implementation}
\hspace*{1em} \textbf{Web agent framework.}
Our system is built on the Magentic-UI web agent framework~\cite{mozannar2025magentic}, which offers a web interface that allows human oversight of web agent actions. 
We chose this web agent framework since its interface displays agent actions on a live webpage with visual overlays (e.g., a rectangle marking the target UI element, a dot showing cursor clicks) so users can monitor execution in real time and spot issues as they arise.
In our implementation, we use Magentic-UI's \textit{web surfer}~\footnote{\url{https://github.com/microsoft/magentic-ui/tree/main/src/magentic_ui/agents/web_surfer}} for web browsing, interaction, and navigation, as well as the same tool definitions (e.g., visit URL, web search, scroll down/up, click, type)~\footnote{\url{https://github.com/microsoft/magentic-ui/tree/main/src/magentic_ui/tools}} to define the web surfer's functionality. 

\textbf{LLMs.}
\tool has been tested with OpenAI GPT-4o~\cite{gpt4o} for webpage parsing and action reasoning of the web surfer. The reasoned actions are performed on the webpage using Playwright~\cite{playwright}. The LLMs for workflow synthesis, condition check on webpage states, and user notification were tested with GPT-4o.

\textbf{Workflow synthesis time.}
The execution time of each task depends on the number of steps. We report the mean and standard deviation of per-task execution time for the formative study tasks in Table~\ref{tab:user_eval_task_execution_time}. To synthesize the \tool Workflow, each original and variation task is executed five times, so the workflow synthesis time is about the total number of executions multiplied by the per-task execution time. To optimize the workflow synthesis time, multiple executions can be initiated in parallel. Failed and successful executions from these runs can then be used for workflow synthesis.

\begin{table*}[!ht]
    \centering
    \footnotesize
    \caption{Execution and workflow synthesis time of user study tasks (mins:secs $\pm$ std).}
    \begin{tabular}{cp{.5\textwidth}cc}
    \toprule
        \textbf{ID} & \textbf{Task Description} & \textbf{Execution} & \textbf{Workflow Synthesis}\\
        \midrule
        1 & Search the <flight type> <cabin type> ticket from <departure city> to <destination city> on <date> for <number of passengers> on <website> & $2{:}38 \pm 0{:}41$ & $\sim52{:}40$\\
        \midrule
        2 & Go to <website> to find <home filtering> <home type> for <rent or buy> in <city>. Scrape the information from top <num listings> & $1{:}20 \pm 0{:}47$ & $\sim26{:}40$\\
        \midrule
        3 & Go to <website> to find the first <num pets> <pet filtering> <pet type>'s <pet information> that can be adopted near <city> & $1{:}33 \pm 0{:}31$ & $\sim31{:}00$\\
        \midrule
        4 & Search the lowest price of <product> with <attributes> by applying <discount> on <website> & $1{:}11 \pm 0{:}38$ & $\sim23{:}40$\\
        \midrule
        5 & Search the <paper title> on Google Scholar. Find the papers authored by <author of interest> of the searched article. Retrieve the most recent paper according to <sorting criteria> and summarize its abstract & $0{:}46 \pm 0{:}20$ & $\sim15{:}20$\\
        \midrule
        6 & Search <news topic> from <website>, and summarize its author's recent articles & $1{:}09 \pm 0{:}39$ & $\sim23{:}00$\\
    \bottomrule
    \end{tabular}
    \Description{This table summarizes the user study tasks, and the time taken to execute the tasks and synthesize workflows.}
    \label{tab:user_eval_task_execution_time}
\end{table*}

%% file: sections/5-auto_evaluation.tex
\section{Benchmark Evaluation} \label{sec:benchmark_evaluation}
To evaluate the effectiveness of workflows synthesized by \tool, we conducted a benchmark evaluation that compares how well different workflows guide web agents to execute repetitive tasks. 

\subsection{Evaluation Setup}
We randomly sampled fifteen tasks from the Skyvern web benchmark~\cite{skyvern} for the benchmark evaluation (task details in Table~\ref{tab:auto_eval_tasks}). For each task, we automatically generated variation tasks (using the method introduced in Section~\ref{sec:auto_workflow_synthesis}), executed the original and variation tasks (a task family) five times, and then synthesized a workflow for each task family. All original tasks had three variations (attribute, category, website) except for Archive (2 variations) and Cars (1 variation), where the missing variations failed in all executions due to captcha errors. 

\begin{figure}[!ht]
    \centering
    \includegraphics[width=.98\linewidth]{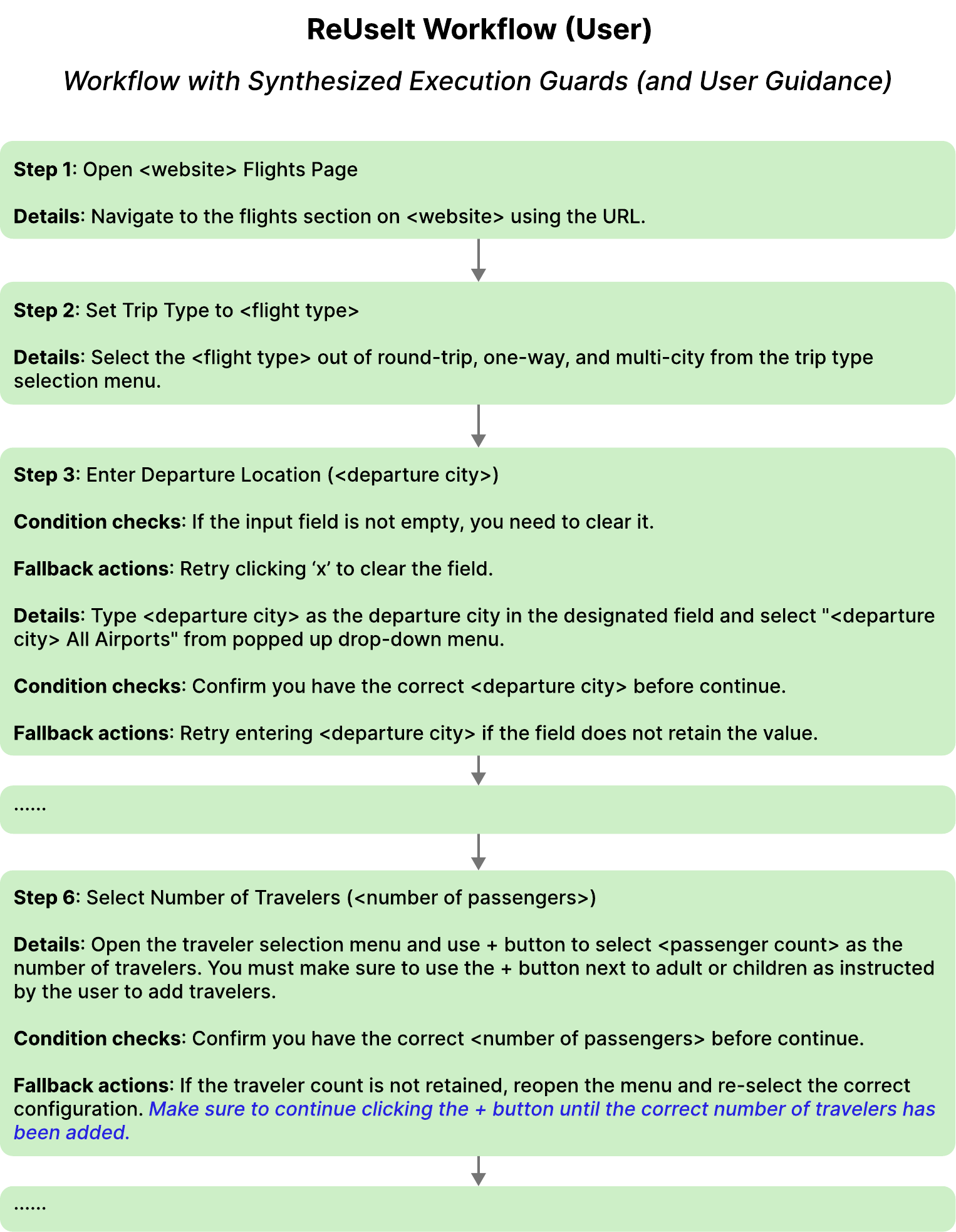}
    \caption{Example of \tool Workflow (User). User guidance used to update the workflow is shown in blue text.}
    \Description{This figure shows an example of "ReUseIt Workflow (User)". The workflow contains the major actions to take for each step, and for each action, it includes the synthesized condition checks and fallback actions. When user feedback is collected, it is added to the workflow to improve the condition checks or fallback actions.}
    \label{fig:workflow_reuseit}
\end{figure}

We compared the agent guided by Task + our synthesized \tool Workflow (e.g., Figure~\ref{fig:workflow_reuseit}) against three baselines: Task-Only, Task + Success-Traces, and Task + Magentic-UI Plan. 
We used the average success rate ($SR$) as the evaluation metric (as introduced in Section~\ref{sec:formative_user_study}). Following WebVoyager~\cite{he2024webvoyager}, we employed LLM-as-a-judge to assess execution success. Specifically, a GPT-4o evaluator takes the task description, the final three screenshots prior to task termination, and the agent's answer, and returns a binary success decision accompanied by a rationale. To validate the performance of this LLM judge, we randomly audited 45 judgments (15\%) with a human rater. The LLM-judge achieved $\textrm{Accuracy} = 0.778$, $\textrm{Precision} = 0.852$, $\textrm{Recall} = 0.793$, $\textrm{F1} = 0.821$. Agreement between the LLM judge and human rater was moderate-substantial with $\textrm{Cohen's }\kappa = 0.528$.

\begin{table*}[!ht]
\centering
\footnotesize
\setlength{\tabcolsep}{4pt}
\renewcommand{\arraystretch}{1.05}
\caption{Web automation tasks in the benchmark evaluation.}
\begin{tabularx}{\linewidth}{@{}p{2cm}Y@{}}
\toprule
\textbf{Task} & \textbf{Details} \\
\midrule
Archive & Search for “Space images” on \texttt{archive.org} and output the capture dates and titles of the first 10 images listed. Only use \url{http://archive.org}. \\
\midrule
Asus & Locate the ASUS support FAQ explaining how to update router firmware and list the step-by-step instructions provided. Only use \url{http://asus.com}. \\
\midrule
Cars & Find detailed specs, including fuel type and VIN, for the lowest-mileage 2020 Toyota Camry offered by a local dealer in Dallas, TX. Use \url{http://cars.com}. \\
\midrule
Bandcamp & Locate an artist’s page (e.g., “Tame Impala”) and list the available album formats (MP3, FLAC, etc.) offered on that page. Only use \url{http://bandcamp.com}. \\
\midrule
BBC & Navigate to the Live section and give the first topic that has “Trump” in it. Use \url{http://bbc.com}. \\
\midrule
Fda & Look up the latest FDA guidance on AI/ML in medical device software and summarize the key points in the introduction. Only use \url{http://fda.gov}. \\
\midrule
Forbes & Search Forbes for the latest startup that raised over \$500 million. Only use \url{http://forbes.com}. \\
\midrule
Gettyimages & Filter “vintage journalism” images by the “Editorial” category and summarize basic info for the first 5 assets. Only use \url{http://gettyimages.com}. \\
\midrule
Groupon & Search spa \& wellness deals in Los Angeles, CA; filter to under \$50; extract merchant ratings for the first five deals. Only use \url{http://groupon.com}. \\
\midrule
Restaurantguru & Filter restaurants in London by price “€€” and cuisine “Sushi,” then list the first 5 result names. Only use \url{http://restaurantguru.com}. \\
\midrule
Scribd & Search “legal case studies,” filter to English and Italiano, and list the top three docs with page count and uploader. Only use \url{http://scribd.com}. \\
\midrule
Smithsonianmag & Browse the Innovation section and extract the top 5 recurring topics/keywords in article summaries; output as a list. Only use \url{http://smithsonianmag.com}. \\
\midrule
Sportskeeda & Find a “Premier League” article and retrieve the section containing expert analysis on a recent match. Only use \url{http://sportskeeda.com}. \\
\midrule
Ticketmaster & Search upcoming concerts in New York City, filter by “Rock,” and list event names, dates, and venues for the next 5 events. Only use \url{http://ticketmaster.com}. \\
\midrule
Tvguide & Analyze the TV Schedule to identify which channel has the most movies scheduled during 6–9 PM. Only use \url{http://tvguide.com}. \\
\bottomrule
\end{tabularx}
\Description{This table presents the web automation tasks.}
\label{tab:auto_eval_tasks}
\end{table*}

\subsection{Evaluation Results}
\begin{figure*}[!ht]
    \centering
    \includegraphics[width=\textwidth]{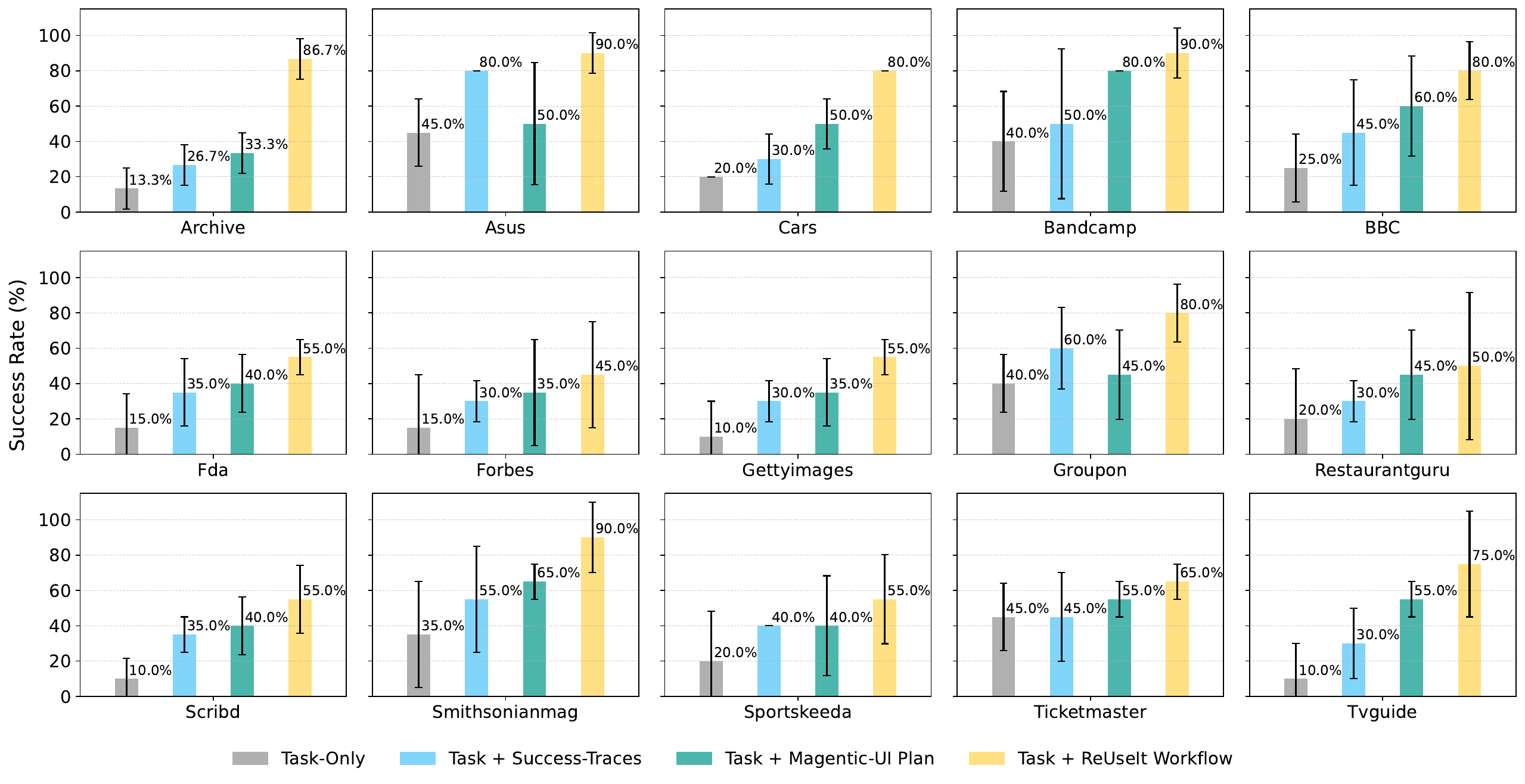}
    \caption{Benchmark evaluation results of fifteen web automation tasks and their variations (avg $\pm$ std success rate).}
    \Description{This figure shows 15 bar plots with task success rates for the benchmark evaluation tasks.}
    \label{fig:task_grid_means}
\end{figure*}

Figure~\ref{fig:task_grid_means} presents the benchmark evaluation results for all fifteen task families. We observe that our synthesized workflow has improved the $SR$ by a large margin compared to the baseline methods. Averaging across tasks, the $SR$s are: Task + \tool Workflow ($70.1\% \pm 16.4\%$), Task + Magentic-UI Plan ($48.6\% \pm 12.9\%$), Task + Success-Traces ($41.4\% \pm 14.8\%$), Task-Only ($24.2\% \pm 13.2\%$). \tool Workflow improves $SR$s by about $21.5\%$ over Task + Magentic-UI Plan, $28.7\%$ over Task + Success-Traces, and $45.9\%$ over Task-Only. Appendix~\ref{sec:benchmark_eval_sr_mean_ci} reports additional results on the mean $SR$s with 95\% confidence intervals for each and across all fifteen task families.

From the results, we find that providing the agent with successful traces has been effective in improving the $SR$s on the original task, but less effective on the variation tasks compared to Magentic-UI Plan. The largest improvements come from adding execution guards in the \tool Workflow. The condition checks and fallback actions have improved agent \textit{reusability} by guiding the agent to reflect on actions and recover from failures. These benefits also boosted \textit{reliability} in executing the same and similar tasks. Additionally, our evaluation spans multiple variation tasks, so higher success rates indicate better \textit{generalizability} to handle value adjustments and webpage UI differences. 

\subsection{Ablation Study}
To further evaluate the effectiveness of components in our synthesized workflows, we conduct an ablation study to evaluate the effectiveness of fallback actions. In this ablation, we implement a variation named {\bf Task + ReUseIt Workflow w/o Fallback Actions}: when condition checks are not met, instead of relying on synthesized fallback actions, the web agent needs to explore recovery strategies on its own and retry the failed step (up to three times). We summarize the ablation results of each and across all benchmark evaluation tasks in Table~\ref{tab:ablation_study}. For most tasks, synthesized workflows with fallback actions achieve higher $SR$s than those without. While simple retries give web agents more attempts after failure, unguided retries are less effective than retries guided by the synthesized fallback actions, which enable agents to recover from failures more reliably. Upon close inspection, we noticed that agents were likely to repeat errors they made in their first attempts when trying to fix errors, which highlights that both condition checks and fallback actions are necessary for improving agents' reliability.

\begin{table}[!ht]
\footnotesize
\centering
\caption{Ablation study results. Mean success rates (\%) $\uparrow$ $\pm$ std of each task and across tasks.}
\Description{This table presents the ablation study results. There are three columns in the table, from left to right, the "Task", the results of "Task + ReUseIt Workflow w/o Fallback Actions", and the results of "Task + ReUseIt Workflow".}
\label{tab:ablation_study}

\begin{tabular}{lcc}
\toprule
\textbf{Task} &
\makecell{\textbf{Task + ReUseIt Workflow} \\ \textbf{w/o Fallback Actions}} &
\textbf{Task + ReUseIt Workflow} \\
\midrule
Archive & $46.7 \pm 11.6$ & $\textbf{86.7} \pm 11.6$ \\
Asus & $60.0 \pm 16.3$ & $\textbf{90.0} \pm 11.6$ \\
Cars & $70.0 \pm 14.1$ & $\textbf{80.0} \pm 0.0$ \\
Bandcamp & $60.0 \pm 0.0$ & $\textbf{90.0} \pm 14.1$ \\
BBC & $45.0 \pm 19.2$ & $\textbf{80.0} \pm 16.3$ \\
Fda & $35.0 \pm 10.0$ & $\textbf{55.0} \pm 10.0$ \\
Forbes & $\textbf{45.0} \pm 10.0$ & $\textbf{45.0} \pm 30.0$ \\
Getty images & $45.0 \pm 19.2$ & $\textbf{55.0} \pm 10.0$ \\
Groupon & $55.0 \pm 10.0$ & $\textbf{80.0} \pm 16.3$ \\
Restaurantguru & $40.0 \pm 28.3$ & $\textbf{50.0} \pm 41.6$ \\
Scribd & $40.0 \pm 28.3$ & $\textbf{55.0} \pm 19.2$ \\
Smithsonianmag & $65.0 \pm 19.2$ & $\textbf{90.0} \pm 20.0$ \\
Sportskeeda & $40.0 \pm 28.3$ & $\textbf{55.0} \pm 25.2$ \\
Ticketmaster & $50.0 \pm 20.0$ & $\textbf{65.0} \pm 10.0$ \\
Tvguide & $55.0 \pm 25.2$ & $\textbf{75.0} \pm 30.0$ \\
\midrule
Across Tasks & $50.1 \pm 10.3$ & $\textbf{70.1} \pm 16.4$ \\
\bottomrule
\end{tabular}

\end{table}

%% file: sections/6-user_evaluation.tex
\section{User Study}
To understand how our system affects the user experience when interacting with web agents, we conducted a user study to explore the following research questions: 

\begin{itemize}[leftmargin=*]
    \item \textbf{S2-RQ1}: How does the \tool Workflow affect the agent performance to execute repetitive tasks? Can the \tool Workflow be effectively reused in related tasks? 
    \item \textbf{S2-RQ2}: How does the \tool Workflow affect users' ability to identify and understand issues? Can users find issues with reduced effort? 
    \item \textbf{S2-RQ3}: How does the \tool Workflow influence the way users provide guidance? Do they offer reduced guidance and feel confident about the usefulness?
    \item \textbf{S2-RQ4}: How does the agent \tool Workflow impact user willingness to adopt web agents for their own tasks? What aspects affect their thoughts? 
\end{itemize}

\subsection{Study Design} \label{sec:user_study_design}
\hspace*{1em} \textbf{Tasks and baseline methods.}
We used the same tasks as the formative study. For each task, we generated two variation tasks, resulting in three tasks in each task family. The three tasks were executed by web agents with Task-Only, Task + Magentic-UI plan, and Task + \tool Workflow, respectively, for a comparison.

\textbf{Participants.}
We recruited nine participants from a large company (4 female, 5 male; mean age = 27.78, SD = 2.68; IDs S2-P1 to S2-P9). Four participants were from the formative study, and they worked on different tasks. From a pre-study questionnaire, all participants used LLMs daily. For writing prompts for simple daily tasks, 88.89\% strongly agreed and 11.11\% agreed that they have experience; for detailed prompting for complex tasks, 66.67\% strongly agreed and 33.33\% agreed; and for extensive prompting that involves designing, refining, and chaining across multi-stage tasks, 55.56\% strongly agreed, 33.33\% agreed, and 11.11\% disagreed.

\textbf{Procedure.}
The onboarding session ($\sim$10 mins), involving consent, introduction, and an example task walkthrough, followed the same procedure as the formative study. 
After it, each participant examined two task families covering six agent executions ($\sim$30 mins). We shuffled the task order and ensured each task family was examined by three different participants. In each task, they were provided with a slide deck containing the task description, the screen recording of agent actions, the user notification message, and the workflow showing a sequence of actions the agent followed. After examining all the information, participants were asked to verbally analyze why and how the agent failed or successfully completed the tasks, evaluate the helpfulness of the user notification message to understand the agent's issues, and (optionally) offer guidance by either writing a short instruction or typing their edits in the workflow. We recorded the desktop screen and transcribed the audio. 
After the task, participants completed a post-study questionnaire and had a semi-structured interview ($\sim$20 mins). The interview questions can be found in Appendix~\ref{appendix:user_eval_interview}. Each participant was paid a \$30 gift card to acknowledge their participation. 

\textbf{Data analysis.}
We performed a follow-up experiment on the \tool Workflow with participant-offered guidance. In addition, we analyzed interview data using thematic analysis to code the data to surface recurrent patterns and organized the resulting codes into themes aligned with our research questions.

\subsection{Quantitative Measures}
Here we report the follow-up experiment results and questionnaire responses to understand S2-RQ1 to S2-RQ3 regarding the \tool Workflow's impact on agent execution reliability, user understanding of agent issues, and required user guidance. 

\begin{figure*}[!ht]
    \centering
    \includegraphics[width=\linewidth]{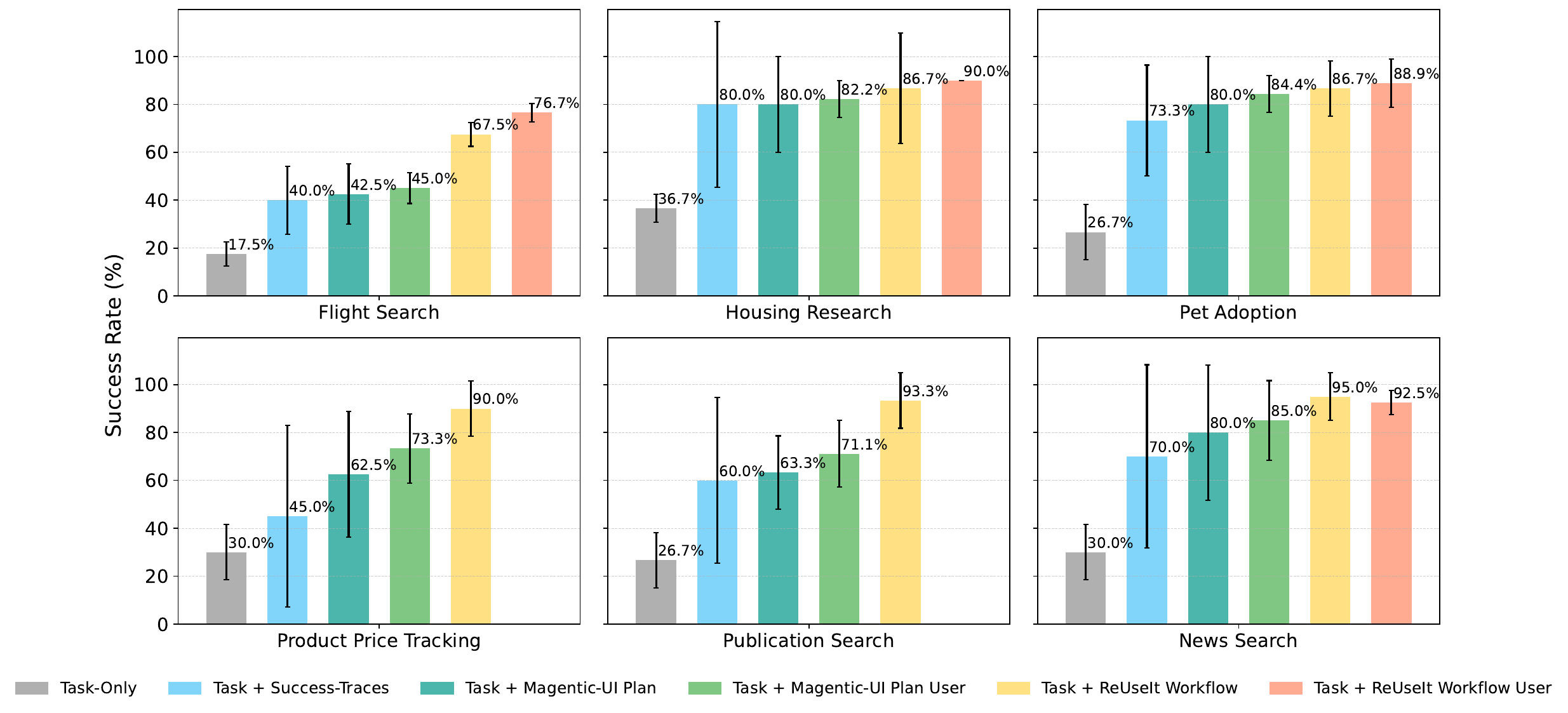}
    \caption{User study follow-up experiment results of six web automation tasks and their variations (avg $\pm$ std success rate). For product price tracking and publication search, users were satisfied with the agent performance in the \tool Workflow condition and did not edit the workflow or offer guidance (i.e., no result of Task + \tool Workflow User is shown).}
    \Description{This figure shows six bar charts with task success rates for the user study follow-up experiment results.}
    \label{fig:user_eval_s2_grid}
\end{figure*}

\textbf{Follow-Up Experiment (S2-RQ1).} \label{sec:user_eval_followup}
Figure~\ref{fig:user_eval_s2_grid} shows the follow-up experiment results and Appendix~\ref{sec:user_study_sr_mean_ci} presents additional results of mean success rates with 95\% confidence intervals for each task and across all tasks. In tasks ``Product Price Tracking'' and ``Publication Search,'' participants were satisfied with the \tool Work\-flow and did not improve it, and we plotted the success rate averaged across only the users who offered guidance in Figure~\ref{fig:user_eval_s2_grid} (\textbf{Task + \tool Workflow User}). See Figure~\ref{fig:workflow_reuseit} for an example workflow updated with user guidance (blue text). 
Appendix~\ref{appendix:userstudy_followup_results_full} reports the full results for each user. Across tasks, the success rates for Task + \tool Workflow and Task + \tool Workflow User were $86.5\% \pm 9.9\%$ and $87.0\% \pm 7.1\%$. These results show that (1) \tool Workflow can boost the agent's task execution reliability compared to the baseline methods and (2) user guidance can further help improve the workflow for better agent execution performance on repetitive tasks. 

\begin{figure*}[!ht]
    \centering
    \includegraphics[width=\textwidth]{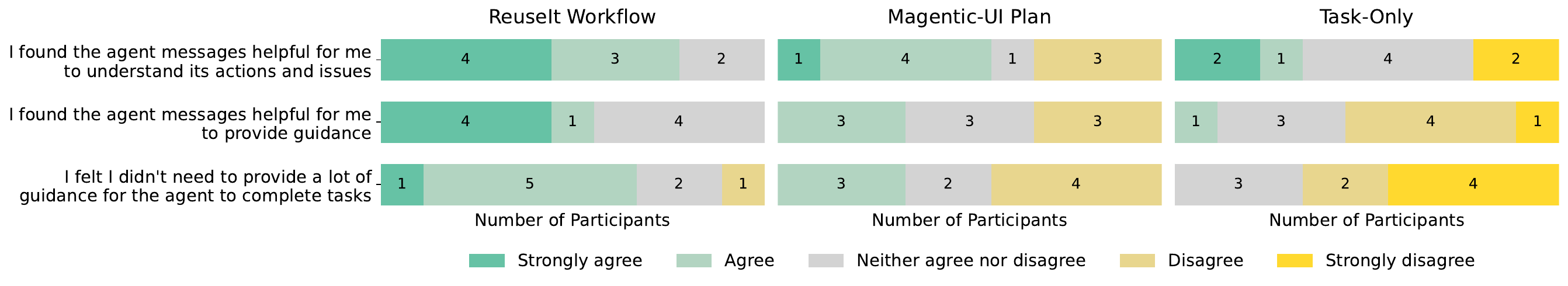}
    \caption{User study post-study questionnaire results. These questions collect user feedback on the agent messages' usefulness to help users understand agent issues and offer guidance, as well as the user-perceived effort on debugging the agent.}
    \Description{This figure presents the responses to the Likert-scale questions from the user study participants. The questions are regarding the agent message's helpfulness to help users understand agent issues and offer guidance, as well as the user-perceived effort on debugging the agent.}
    \label{fig:likert_stacked_bar}
\end{figure*}

\textbf{Questionnaire (S2-RQ2, S2-RQ3).}
Figure~\ref{fig:likert_stacked_bar} shows the post-study questionnaire results. 
Aggregating the three questions per condition, \tool Workflow drew 66.7\% positive and 3.7\% negative responses, with a mean Likert score of 0.96 on a -2 to +2 scale. 
Magentic-UI Plan received 40.7\% positive and 37.0\% negative responses, with a mean of 0.07, essentially neutral overall. 
Task-Only had 14.8\% positive vs. 48.1\% negative responses, mean -0.52. 
By question, \tool Workflow helped to provide user notification messages that were the most helpful to understand agent actions/issues (7 positive, 2 neutral) and provide guidance (5 positive, 4 neutral), reducing the need for user guidance (6 positive, 2 neutral, 1 negative). 
Magentic-UI Plan helped some participants understand actions/issues (5 positive, 1 neutral, 3 negative), but received divided opinions on user guidance (3/3/3 split) and leaned negative on reduced guidance (3 positive, 2 neutral, 4 negative). 
Task-Only was weakest on all fronts, especially helpfulness to provide guidance (1 positive, 3 neutral, 5 negative), and little needed guidance (3 neutral, 6 negative). 

Accompanying the question on required user guidance, we conducted an analysis of user guidance statistics to compare the number and length of user guidance. We received 9, 17, and 17 pieces of user guidance to \tool Workflow, Magentic-UI Plan, and Task-Only, respectively. The number of words (mean $\pm$ std) in user guidance was $43.41 \pm 50.02$ (\tool Workflow), $43.24 \pm 48.18$ (Magentic-UI Plan), and $56.11 \pm 44.59$ (Task-Only). 
These results indicate that the execution guards in the \tool Workflow have effectively improved the interpretability of agent actions/issues and reduced the guidance burden on users. 

\subsection{Qualitative Findings} 
From the analysis of interview data, we obtain the following findings centered around S2-RQ2 to S2-RQ4. Overall, we found that the \tool's synthesized workflows have helped users to spot and fix agent issues and augmented their willingness to adopt web agents. 

\textbf{Condition checks helped users to identify issues, issue analysis helped them to understand challenges (S2-RQ2).}
Participants emphasized that \tool Workflow's condition checks helped them anticipate and localize failures without exhaustive monitoring. S2-P3 noted, \textit{``If you know where the model's gonna fail from the condition checks [in \tool Workflow], you don't have to watch the agent very carefully and examine everything. You kind of know when the error's gonna be. Otherwise, there's always some possibility of making an error at any step. So definitely helps you pinpoint where things are gonna go wrong.''}
They also commented that the user notification guided by \tool Workflow clarified the reason why the agent failed, which made issue diagnosis easier. 
For instance, S2-P4 said that \textit{``With \tool Workflow, I would understand a little bit better why,''} and S2-P3 emphasized the value of this detail: \textit{``For \tool Workflow, I feel like even if it gets it wrong, it gives you so much detail that you can exactly intervene on the spot that you want to.''}

By contrast, user notification with Magentic-UI Plan and Task-Only sometimes failed to identify the agent issues and lacked an analysis of the agent actions when it failed (S2-P1, S2-P2, S2-P3). 
As S2-P2 put it, \textit{``I feel like oftentimes it [the agent] just sad I was successful, but sometimes it's just not really successful. I think if it turns out that it becomes unsuccessful, the message is probably not gonna be very informative for me to know what's happening.''}

\textbf{User notification with challenge identification and guidance tips allowed precise and low-effort user guidance (S2-RQ3).}
Participants described the user notification provided with \tool Workflow broadly covered the agent's challenges that supported targeted edits. As S2-P1 explained that \textit{``The messages are like a super set of the challenges faced by the agent. I always find that my instruction is a subset of the suggested ideas. And also, I think it's important that I match the suggested ideas with the actual workflow of the agent because some of the suggested content is already covered by the workflow, so I don't have to repeat it. Instead, I need to think about why the existing workflow is not good enough.''} 
In addition, S2-P2 underscored the usefulness of actionable guidance, saying that \textit{``I do think that the actionable guidance says something about potentially what are some of the next steps, and after reviewing the workflow myself and then editing the prompts, I think the actionable guidance does have a good contribution.''}
The \tool Workflow, in turn, supported user understanding of agent challenges for them to provide more effective guidance (S2-P1, S2-P8). For example, S2-P8 said that \textit{``These [condition checks and fallback actions] actually give me not only hints, but also much guidance on how I should train an agent, or what my approach should be.''}

However, user notification with Magentic-UI Plan and Task-Only sometimes increased user effort and caused uncertainty when offering guidance. S2-P5 contrasted the experiences, saying, \textit{``I guess with \tool Workflow, they are usually very comprehensive and correct. So it's much easier for me to provide guidance. It's basically just confirming what I saw and then translating it into updating those steps. The Task-Only and Magentic-UI Plan is basically, I have no idea what went wrong or if it actually went wrong, so I can only be specific about what I want and hope that it understands the problem, understands the UI, and executes it.''}

\textbf{Reliable and controllable web agents improved user willingness for practical use (S2-RQ4).}
Participants highlighted that the \tool's workflows made agent reuse feasible because the agent behaved more consistently and could be steered through explicit condition checks. As S2-P2 put it, \textit{``If I wanna reuse it [the agent] reliably, then I would consider being very extensive with my workflow and being very consistent about it... so that I know how it's supposed to work.''} S2-P4 further commented on the connection between user trust and agent reliability, saying, \textit{``I think I have a better chance of getting better results just because it [the agent] has... ways of approaching the task, so I think I would trust it more. Especially if I had a certain level of control in terms of the conditions and the fallbacks, I would use it even more just to make sure that the result that I'm getting is the proper one.''}

%% file: sections/7-discussion.tex
\section{Discussion} \label{sec:discussion}
Based on our evaluation findings, here we discuss the implications for design, opportunities with \tool, current limitations, and future directions.

\subsection{Implications for Design} \label{sec:design_implications}
\hspace*{1em} \textbf{Accurate issue detection and explanation lay the foundation for user understanding and trust.}
Our user study results reveal that accurate agent error identification is critical for user understanding and trust of web agents. If the agent misreports its own behavior, user trust can collapse (S2-P2, S2-P5).
This finding aligns with broader research on AI transparency and trust, where users need to monitor and understand the AI's workings in order to trust it~\cite{kassick2024fostering}. When an AI system conceals errors or provides only vague error information, users cannot calibrate their trust appropriately and may become frustrated~\cite{kassick2024fostering, berkowitz2025when}. 

To boost user trust, systems can incorporate verification mechanisms that confirm an agent's claims against reality by performing explicit action checks using multiple signals, such as document object model (DOM) state changes, visual screenshots, and console logs. This type of cross-validation is related to the verifier component in advanced GUI agents, which inspects the environment before and after each action to evaluate whether the result matches the agent's expectation~\cite{tang2025survey}. Along with this, intelligent error discrimination has been explored to filter noise, such as distinguishing transient or benign issues (e.g., a momentary delay in loading) from true failures, suppressing false alarms~\cite{jackson2025agentic}. We also observed the LLM judge tending to produce false negatives in our experiments (Section~\ref{sec:user_eval_followup}), and these verification techniques can be helpful to ensure a more reliable error detection. 

\textbf{Hierarchical explanations and progressive disclosure of agent messages for better user understanding.}
Participants in our study emphasized the need for hierarchical explanations to improve the web agent's interpretability. They wanted the agent's message delivery to start with a concise top-line summary, followed by more detailed reasoning revealed only on demand (S2-P7). This layered format aligns with the UX principle of progressive disclosure, which suggests showing only essential information first and uncovering specificity progressively~\cite{nielsen2006progressive}. Such progressive disclosure can help users grasp the outcome without being overwhelmed~\cite{nielsen2006progressive, mueller2021principles, kreiterling2025design}. 
Furthermore, one participant (S2-P9) specifically endorsed that the agent's response should clearly identify the problem, explain the cause, and provide guidance on solutions. This format is supported by user-interface guidelines on effective messaging in UX design~\cite{microsoft2025uxforerrors, minhas2024how}, and can help future research to structure agent messaging in human-agent interaction.

\textbf{Reveal the process of agent decision-making for user debugging and trust.}
Our participants suggested that visibility into the agent's decision process can be a helpful strategy to debug failures and build trust. This means exposing why the agent took a certain action and what it considered along the way~\cite{eriksen2025designing}. They specifically mentioned that a useful system can reveal the agent's consideration set over time, e.g., what the agent looked at or weighed before acting (S2-P1). This insight aligns with developments in XAI that visualize AI's attention~\cite{an2022attention}. Translating this to web agents suggests that we could visually indicate which UI elements the agent considered at each step by highlighting them. In this way, users can understand how an action was proposed and trace the cause of a failure, such as overlooking a critical button or misprioritizing a less relevant link. 

Additionally, another interesting insight was revealing the agent's inactions (S2-P5). This idea resonates with the concept of contrastive understanding in XAI, which deals with ``why'', ``why not'' questions, and hypothetical ``what if'' scenarios by comparing the actual outcome with alternatives~\cite{bao2024understanding}. Prior HCI research has demonstrated the value of this approach by showing explanations of why an action was taken and why others were not can significantly improve system understandability~\cite{krarup2019towards}. Thus, a design that surfaces the UI elements evaluated and explicitly states the detailed rationale can help to build more interpretable web agents. 

\textbf{Correlate text messages with interface actions for verification.}
Participants in our study requested an auto-transcriber (S2-P4) and simultaneous commentary while the interface is updating (S2-P7) for easy agent behavior understanding. Prior HCI literature echoed that agents need to explain their actions while acting to allow easy inspection~\cite{lieberman2003agents}. This lets users catch mistakes and understand the outcome of each action, which is crucial for user acceptance (e.g., justifying decisions, ensuring safety)~\cite{lieberman2003agents, rosenfeld2019explainability}. 
In fact, recent UX guidelines for AI recommend showing live status indicators as side-panel histories of every API call or action, because exposing the AI's chain of thought reduces user anxiety about black-box behavior by around 34\%~\cite{zhu2025designing}. 

Beyond simply describing actions, it is vital that claims are verifiable against the visuals. Our users desired to see confirmation of what the agent said it did (S2-P5). This matches findings from voice-assistant studies in which users trust results more with visible evidence~\cite{budiu2018intelligent}. 
Likewise, if the web agent says ``I scrolled to the bottom of the page,'' the interface should visibly highlight the bottom or timestamp that action so users can confirm it. Concretely, every narrated step could tag the time in the activity log and outline the relevant UI element. If the agent's summary does not match the actual UI state, the system can flag a discrepancy so the user notices. 

\subsection{Opportunities in \tool}
\hspace*{1em} \textbf{Scale to broader web automation tasks.}
To extend \tool for broader tasks, each unit in the synthesized workflow can be treated as a reusable building block. Inspired by Retrieval-Augmented Generation (RAG)~\cite{lewis2020retrieval}, we can maintain a library of workflow units, each consisting of an action, its condition checks, and fallback actions, and create new workflows by retrieving the relevant units and assembling them. In practice, units can be indexed with metadata (e.g., task, website, UI pattern, DOM, screenshots) so that future tasks can retrieve, adapt, and connect functionally relevant units. This RAG-based approach can help to scale coverage across diverse web tasks by reusing workflow units. 

\textbf{Tailor user notification based on user and task.}
The right content and style of user notification should depend on both the user and the task. Our participants commented that for simple tasks (e.g., price search, news search), they preferred reliability over interpretability (S2-P3, S2-P7). 
For complex tasks (e.g., reading papers, aggregating statistics), on the other hand, our participants desired fine-grained workflows to dive deeply into how each action is taken for rigorous user review (S2-P3). 
These insights suggest valuable research directions for tailoring agent messaging and control according to user needs. We can surface historical success rates via public data (e.g., a leaderboard), so users can quickly gauge how likely their task may succeed. Based on this information, users can have the option to choose an \textit{assurance} mode to allow the agent to run in the background and return a compact outcome, or a \textit{research} mode with fine-grained traces like action logs, UI elements considered, and exportable artifacts for detailed review. 

\textbf{Extend the workflow synthesis approach to support other types of AI agents.}
Our work introduces a workflow synthesis approach to augment the reliability and reusability of web agents. The idea of synthesizing execution guards to guide agent reflection and recovery applies to broader AI agents, such as GUI agents (e.g., agents in mobile UIs~\cite{wang2024mobile} and computer operating systems~\cite{wu2024copilot, xie2024osworld}) and API-based AI agents~\cite{shen2024shortcutsbench}. The workflow synthesis can follow a similar trial-and-error process to derive condition checks and fallback actions. To construct condition checks, GUI agents can use screenshots or system states, and API agents can verify the API call status and responses. Besides fully automated approaches, we may design human-in-the-loop approaches where users and agents can examine and correct suspicious or erroneous states together. In these applications, we can still leverage issues surfaced by failed condition checks to notify users to help them monitor and debug agent workflows so they do not need to manually track the progress, which can be especially helpful when lower-level action traces are difficult to interpret (e.g., API agents).

\subsection{Limitations and Future Work}
\hspace*{1em} \textbf{Level of workflow details, and reusability vs. generalizability tradeoff.}
Our current workflow synthesis has been affected by the web agent capability, such that extensive condition checks were synthesized from failed executions to improve web agent reusability. In the meantime, it raises a question regarding the reusability vs. generalizability tradeoff. 
Intuitively, workflows with fewer restrictions can afford better generalizability as the agent is expected to adapt its actions in varying scenarios. However, our experiments show that agents following fewer restrictions (e.g., Magentic-UI Plan) had more random actions, which downgraded their reliability to reasonably generalize. This means that current web agents still require specific and detailed instructions to achieve the expected behavior, which can lead to verbose workflows and increase the user's burden to offer guidance when the agent fails to follow the workflows.
We envision that as agents evolve with improved planning and execution capabilities, they can adhere more strictly to planned actions and exhibit less random behavior. In such scenarios, tightened workflows can be synthesized to guide agents. This would reduce the time required to synthesize condition checks and fallback actions from trial-and-error and minimize the need for user oversight and intervention to recover from low-level but critical mistakes. Instead, high-level workflows consisting of major steps could be employed, allowing user involvement to focus on personalizing workflows for their specific needs. With this approach, future work could further explore the tradeoff between reusability and generalizability, balancing the level of details to satisfy user needs for either execution accuracy or task extendability.

\textbf{User intervention with agent execution checkpointing and branching.}
\tool allows human-in-the-loop intervention at points where the agent fails after retries, which forms a linear way of user intervention where the agent pauses the action, requests user guidance, and resumes the execution. In addition to this interaction method, there are cases where users may need to intervene at a previous step because an earlier prerequisite is not satisfied, e.g., the agent cannot submit a form because a required entry has not been filled out yet. To support this type of user intervention, future work can explore using agent frameworks like LangGraph~\cite{langgraph} or debug tools like AGDebugger~\cite{epperson2025interactive} for automatic checkpointing. These techniques can save the execution state after each action so the agent or user can rewind to that point, update instructions, and fork a new execution. 
Based on the idea of checkpointing, future research can also study workflow branching to explore alternative execution paths. Prior work has found that Directed Acyclic Graph (DAG)-based workflows with parallel branches can significantly boost the efficiency of LLM agents~\cite{qiao2024benchmarking}. With human-in-the-loop agent execution, automatically synthesized workflows can benefit from human experience and knowledge to include branches and propagate that to future workflow synthesis for efficient task execution.

%% file: sections/8-conclusion.tex
\section{Conclusion}
End users and professionals often need to execute web tasks repetitively. Although AI-powered web agents have the potential to automate repetitive tasks, they lack the reliability to execute these tasks without human guidance. We present \tool, an automatic approach to synthesize workflows with execution guards that can guide web agents during task execution. The execution guards incorporate condition checks and fallback actions that allow agents to identify and recover from errors for higher execution reliability, as well as help users to understand issues for better agent interpretability. In a benchmark evaluation on fifteen tasks, we found \tool's synthesized workflows increased the average success rate from 24.2\% to 70.1\%. In a user study with nine users, we observed that the synthesized workflows helped users to easily understand agent failures and boosted their trust and willingness to adopt web agents.

%% file: sections/appendix.tex
\onecolumn
\clearpage

\section{Preliminary Evaluation Tasks and Results} 
This section presents the tasks used in the preliminary evaluation and formative study, and the preliminary evaluation results (Section \ref{sec:prelim_eval}). 

\subsection{Tasks} \label{appendix:formative_tasks}
Table \ref{tab:user_eval_tasks} summarizes the tasks used in the preliminary evaluation and formative study.

\begin{table}[!ht]
    \centering
    \footnotesize
    \caption{Preliminary evaluation and formative study web automation tasks.}
    \begin{tabular}{cp{.45\textwidth}p{.26\textwidth}l}
    \toprule
        \textbf{ID} & \textbf{Task Description} & \textbf{Agent Actions} & \textbf{Task Domain}\\
        \midrule
        1 & Search the <flight type> <cabin type> ticket from <departure city> to <destination city> on <date> for <number of passengers> on <website> & Web navigation, form filling, filtering & \textbf{Flight} Search\\
        \midrule
        2 & Go to <website> to find <home filtering> <home type> for <rent or buy> in <city>. Scrape the information from top <num listings> & Web navigation, filtering, sorting, information extraction & \textbf{Housing} Research\\
        \midrule
        3 & Go to <website> to find the first <num pets> <pet filtering> <pet type>'s <pet information> that can be adopted near <city> & Web navigation, filtering, information extraction & \textbf{Pet} Adoption\\
        \midrule
        4 & Search the lowest price of <product> with <attributes> by applying <discount> on <website> & Web navigation, product search, filtering, price comparison & Product \textbf{Price} Tracking\\
        \midrule
        5 & Search <paper title> on Google Scholar. Find the papers authored by <author of interest> of the searched article. Retrieve the most recent paper according to <sorting criteria> and summarize its abstract & Keyword search, author navigation, sorting, content summarization & \textbf{Publication} Search\\
        \midrule
        6 & Search <news topic> from <website>, and summarize its author's recent articles & Keyword search, author navigation, content summarization & \textbf{News} Search\\
    \bottomrule
    \end{tabular}
    \label{tab:user_eval_tasks}
\end{table}

\subsection{Preliminary Evaluation Results} \label{sec:prelim_eval_sr_mean_ci}
Table \ref{tab:prelim_eval} summarizes the mean success rates (with 95\% confidence intervals) of each task and across all tasks for each condition in the preliminary evaluation. 

\begin{table}[!ht]
\centering
\footnotesize
\caption{Preliminary evaluation results. Mean success rates (\%) $\uparrow$ with 95\% confidence intervals of each task and across tasks.}

\begin{tabular}{@{}c c@{}}
% ================= LEFT BLOCK (TWO STACKED TABLES) =================
\begin{minipage}[t]{0.78\linewidth}
\centering

% ----------- Part 1: First three tasks -----------
\begin{tabular}{lp{0.18\textwidth}p{0.18\textwidth}p{0.18\textwidth}}
\toprule
 &
\textbf{Flight} &
\textbf{Housing} &
\textbf{Pet} \\
\midrule
Task-Only &
17.5 [9.5, 25.5] &
36.7 [22.3, 51.0] &
26.7 [-2.0, 55.4] \\

Task + Success-Traces &
40.0 [17.5, 62.5] &
80.0 [-6.1, 166.1] &
73.3 [16.0, 130.7] \\

Task + Magentic-UI Plan &
42.5 [22.5, 62.5] &
80.0 [30.3, 129.7] &
80.0 [30.3, 129.7] \\

Task + Magentic-UI Plan User &
\textbf{45.0} [34.8, 55.2] &
\textbf{82.2} [63.1, 101.3] &
\textbf{84.4} [65.3, 103.6] \\
\bottomrule
\end{tabular}

\vspace{0.6em}

% ----------- Part 2: Next three tasks -----------
\begin{tabular}{lp{0.18\textwidth}p{0.18\textwidth}p{0.18\textwidth}}
\toprule
 &
\textbf{Price} &
\textbf{Publication} &
\textbf{News} \\
\midrule
Task-Only &
30.0 [11.6, 48.4] &
26.7 [-2.0, 55.4] &
30.0 [11.6, 48.4] \\

Task + Success-Traces &
45.0 [-15.2, 105.2] &
60.0 [-26.1, 146.1] &
70.0 [9.1, 130.9] \\

Task + Magentic-UI Plan &
62.5 [20.7, 104.3] &
63.3 [25.4, 101.3] &
80.0 [35.0, 125.0] \\

Task + Magentic-UI Plan User &
\textbf{73.3} [50.4, 96.2] &
\textbf{71.1} [36.6, 105.6] &
\textbf{85.0} [58.5, 111.5] \\
\bottomrule
\end{tabular}

\end{minipage}
&
% ================= RIGHT BLOCK (ACROSS TASKS) =================
\begin{minipage}[t]{0.20\linewidth}
\centering

\begin{tabular}{lc}
\toprule
\textbf{Across Tasks} \\
\midrule
27.9 [21.3, 34.5] \\
61.4 [44.5, 78.3] \\
68.1 [52.2, 83.9] \\
\textbf{73.5} [57.6, 89.4] \\
\bottomrule
\end{tabular}

\end{minipage}

\end{tabular}
\label{tab:prelim_eval}
\end{table}

\section{Formative Study Materials and Results} 
This section offers the materials (interview questions) used in the formative study and additional results in the follow-up experiments (Section~\ref{sec:formative_user_study}).

\subsection{Interview Questions} \label{appendix:formative_interview}
\subsubsection*{Understanding of Agent Behavior}
\begin{enumerate}
    \item How did the screen recordings help you understand the agent’s behavior?
    \begin{itemize}
        \item Were you able to figure out what the agent did successfully or unsuccessfully based on the provided information? Why?
    \end{itemize}
    \item How did the agent workflow help you understand the agent’s behavior?
    \begin{itemize}
        \item Were you able to figure out what the agent did successfully or unsuccessfully based on the provided information? Why?
    \end{itemize}
\end{enumerate}

\subsubsection*{Workflow Improvement Strategies}
\begin{enumerate}[resume]
    \item When improving the workflow, what did you focus on?
    \begin{itemize}
        \item Why did you focus on these aspects?
    \end{itemize}
    \item What specific strategies did you use to make the workflow more reusable for the same and similar tasks?
    \begin{itemize}
        \item How did you expect these strategies to help in future executions?
    \end{itemize}
    \item Were there any challenges in improving the workflow? If yes, what were they?
\end{enumerate}

\subsubsection*{Using Agents Yourself}
\begin{enumerate}[resume]
    \item Now that you know potential errors or limitations of web agents, what criteria do you think the agent needs to meet to be used in your work/life?
    \begin{itemize}
        \item Consistency
        \begin{enumerate}
            \item Replay of the same task
            \item Variations of similar tasks
            \item Generalize to different tasks
        \end{enumerate}
        \item Tradeoffs between task complexity and agent reliability, flexibility
        \begin{enumerate}
            \item For what types of tasks do you prefer reliability?
            \item For what types of tasks do you prefer flexibility?
        \end{enumerate}
    \end{itemize}
    \item What tasks do you want to put effort into improving the workflow for, and which ones would you rather do yourself?
\end{enumerate}

\subsubsection*{Suggestions for Future Design}
\begin{enumerate}[resume]
    \item What other features or information would make it easier for you to understand agent behaviors and provide guidance in the future?
    \begin{itemize}
        \item Can you elaborate on why you think it would be helpful?
    \end{itemize}
    \item Is there anything else you’d like to share about your experience in this study?
\end{enumerate}

\subsection{Follow-Up Experiment Results} \label{appendix:formative_followup_results_full}
Table \ref{tab:user_task_success} shows the follow-up experiment results for each user in the formative study. 

\begin{table}[!ht]
\centering
\footnotesize
\caption{Formative study follow-up experiment per-user results. Mean success rates (\%) $\pm$ std.}
\begin{tabularx}{0.98\linewidth}{@{}c | *{6}{Y} | l@{}}
\toprule
 & \textbf{Flight} & \textbf{Housing} & \textbf{Pet} & \textbf{Price} & \textbf{Publication} & \textbf{News} & \textbf{Across Tasks} \\
\hline
\textbf{S1-P1} & 45.0 $\pm$ 10.0 & -- & -- & 85.0 $\pm$ 10.0 & -- & 95.0 $\pm$ 10.0 & 75.0 $\pm$ 26.5 \\
\textbf{S1-P2} & -- & 73.3 $\pm$ 11.5 & 73.3 $\pm$ 11.5 & -- & 60.0 $\pm$ 20.0 & -- & 68.9 $\pm$ 7.7 \\
\textbf{S1-P3} & 40.0 $\pm$ 0.0 & -- & -- & 60.0 $\pm$ 16.3 & -- & 75.0 $\pm$ 25.2 & 58.3 $\pm$ 17.6 \\
\textbf{S1-P4} & -- & 86.7 $\pm$ 11.5 & 86.7 $\pm$ 11.5 & -- & 66.7 $\pm$ 11.5 & -- & 80.0 $\pm$ 11.5 \\
\textbf{S1-P5} & 50.0 $\pm$ 20.0 & -- & -- & 75.0 $\pm$ 19.1 & -- & 85.0 $\pm$ 19.1 & 70.0 $\pm$ 18.0 \\
\textbf{S1-P6} & -- & 86.7 $\pm$ 11.5 & 93.3 $\pm$ 11.5 & -- & 86.7 $\pm$ 11.5 & -- & 88.9 $\pm$ 3.8 \\
\bottomrule
\end{tabularx}
\label{tab:user_task_success}
\end{table}

\section{\tool Prompts}
This section offers the prompts used in \tool to generate variation tasks and synthesize condition checks, fallback actions, and workflows (Section \ref{sec:system_design}). 

\subsection{Variation Task Generation} \label{sec:prompt_var_task_gen}
\begin{llmscriptblock}{}
\noindent You are an expert specializing in generating task variations for web agents. Your goal is to analyze a given user task and produce different task variants according to the given requirements.
\vspace{1em}

\noindent\textbf{TASK TO ANALYZE:}
\par\noindent
\llmplaceholder{ORIGINAL TASK}
\vspace{1em}

\noindent\textbf{METHOD:}
\par\noindent
Analyze the input task and identify its variable components (e.g., form values, categories, websites). Based on this analysis, generate three distinct task variations, each corresponding to a different type of variation, while preserving the original task's objective and intent. The three required variation types are:

\begin{itemize}[label=\textendash, wide, nosep, leftmargin=1.5em]
    \item \textbf{Attribute Variation}: Modify specific input values (e.g., dates, quantities, names, locations, or other form fields) that would be entered on the same webpage.
    \item \textbf{Category Variation}: Modify a high-level option that requires switching a tab, toggle, or category within the same website (e.g., changing ``One-way'' to ``Round-trip'' on a flight search site).
    \item \textbf{Website Variation}: Change the target website to a different one while keeping the underlying task objective unchanged (e.g., switching from Expedia.com to Google Flights for searching flight tickets).
\end{itemize}
\vspace{1em}

\noindent\textbf{OUTPUT REQUIREMENTS:}
\par\noindent
Return the output using the exact structure specified below. The original task must be reproduced verbatim, without any modification. Each generated variation must clearly correspond to its designated variation type and remain realistic and actionable for a web agent.
\vspace{0.5em}

\noindent
The output must strictly follow the EXAMPLE OUTPUT FORMAT.
\vspace{1em}

\noindent\textbf{EXAMPLE OUTPUT FORMAT:}
\par
\begin{itemize}[label=\textendash, wide, nosep, leftmargin=1.5em, topsep=2pt]
    \item \textbf{Original Task}: Exact original task text.
    \item \textbf{Attribute Variation Task}: Task with modified attribute values on the same webpage.
    \item \textbf{Category Variation Task}: Task involving a different tab, toggle, or category on the same website.
    \item \textbf{Website Variation Task}: Task performed on a different website with the same objective as the original task.
\end{itemize}

\end{llmscriptblock}

\subsection{Condition Check Synthesis} \label{sec:prompt_condition_check}
\begin{llmscriptblock}{}
\noindent You are an expert specializing in diagnosing failures and formalizing condition checks from these failures for web agents. Your goal is to analyze failed execution messages and convert them into explicit condition checks that should be met to prevent future failures.
\vspace{1em}

\noindent\textbf{INPUT TO ANALYZE:}
\par\noindent
\llmplaceholder{FAILED EXECUTION MESSAGES}
\vspace{1em}

\noindent\textbf{METHOD:}
\par\noindent
Analyze each failed execution message to determine the underlying reason an action did not succeed (e.g., inactive buttons, missing input fields). Identify cues such as ``failed to,'', ``didn't,'', or ``couldn't'' to locate failure points. For each failure, abstract the cause into a clear, verifiable condition that should be checked before or after the corresponding action. 
\vspace{0.5em}

\par\noindent
\textbf{Important Constraint:}
\par\noindent
When deriving condition checks from a failed action, do not include any concrete or literal values from the original action (e.g., specific text strings, numbers, dates, names, or URLs). Conditions must be written using generic, value-agnostic wording that captures the underlying requirement (e.g., element state, page readiness, input availability) rather than the specific instance that caused the failure. The purpose of each condition is to prevent the same class of failure across different inputs or contexts, not to guard against a single, fixed value from the original action.
\vspace{1em}

\noindent\textbf{OUTPUT REQUIREMENTS:}
\par\noindent
For every identified failure, generate a rule using the exact syntax specified below. Each rule must clearly state:
\begin{itemize}[label=\textendash, wide, nosep, leftmargin=1.5em]
    \item The relevant action.
    \item Whether the check should occur before or after the action.
    \item The condition that should be satisfied to avoid the failure.
\end{itemize}
\vspace{0.5em}

\noindent The output must strictly follow the EXAMPLE OUTPUT FORMAT.
\vspace{1em}

\noindent\textbf{EXAMPLE OUTPUT FORMAT:}
\par
\begin{itemize}[label=\textendash, wide, nosep, leftmargin=1.5em, topsep=2pt]
    \item \textbf{Action}: Concise description of the failed action.
    \item \textbf{Condition Check}: Before or after performing \llmplaceholder{Action}, ensure \llmplaceholder{Condition} is satisfied.
\end{itemize}

\end{llmscriptblock}

\subsection{Fallback Action Synthesis} \label{sec:prompt_fallback_action}
\begin{llmscriptblock}{}
\noindent You are an expert specializing in synthesizing retry strategies for web agents. Your goal is to analyze failed actions alongside successful executions and derive fallback actions that can help web agents recover from failures.
\vspace{1em}

\noindent\textbf{INPUT TO ANALYZE:}
\par\noindent
\textbf{Failed Action Description:}
\par\noindent
\llmplaceholder{FAILED ACTION}
\vspace{0.5em}

\noindent\textbf{Successful Execution Messages:}
\par\noindent
\llmplaceholder{SUCCESSFUL EXECUTION MESSAGES}
\vspace{1em}

\noindent\textbf{METHOD:}
\par\noindent
Compare the failed step with the actions described in the successful execution messages of the same action. Identify the specific UI interactions (e.g., clicking, navigating, typing) that enabled success in those executions. Abstract these actions into a clear, specific retry instruction that can be applied to recover from the failure.
\vspace{0.5em}

\par\noindent
\textbf{Important Constraint:}
\par\noindent
When deriving fallback actions from a successful action, do not include any concrete or literal values from the original action (e.g., specific text strings, numbers, dates, names, or URLs). Fallback actions must be written using generic, value-agnostic wording that abstracts the underlying recovery strategy (e.g., satisfy element state, ensure page readiness, check input availability), rather than reproducing a specific successful instance. The purpose of each fallback action is to help web agents recover from the same class of failure across varying inputs, pages, or contexts, not limited to addressing a single, fixed failed action.
\vspace{1em}

\noindent\textbf{OUTPUT REQUIREMENTS:}
\par\noindent
Generate recovery strategies using the exact syntax specified below. The strategies should describe the retry actions in terms of the concrete UI interactions that led to success.
\vspace{0.5em}

\noindent The output must strictly follow the EXAMPLE OUTPUT FORMAT.
\vspace{1em}

\noindent\textbf{EXAMPLE OUTPUT FORMAT:}
\par
\begin{itemize}[label=\textendash, wide, nosep, leftmargin=1.5em, topsep=2pt]
    \item \textbf{Action}: Concise description of the failed action.
    \item \textbf{Fallback Action}: Retry \llmplaceholder{Action} by performing \llmplaceholder{Fallback Action}.
\end{itemize}

\end{llmscriptblock}

\subsection{Workflow Synthesis} \label{sec:prompt_workflow_syn}
\begin{llmscriptblock}{}
\noindent You are an expert specializing in constructing generalizable workflows for web agents. Your goal is to synthesize an executable workflow by integrating condition checks and fallback actions into a given workflow structure.
\vspace{1em}

\noindent\textbf{INPUT TO ANALYZE:}
\par\noindent
\textbf{Workflow Structure:}
\par\noindent
\llmplaceholder{WORKFLOW STRUCTURE}
\vspace{0.5em}

\noindent\textbf{Condition Checks:}
\par\noindent
\llmplaceholder{CONDITION CHECKS}
\vspace{0.5em}

\noindent\textbf{Fallback Actions:}
\par\noindent
\llmplaceholder{FALLBACK ACTIONS}
\vspace{1em}

\noindent\textbf{METHOD:}
\par\noindent
For each relevant step in the given workflow structure, insert the corresponding pre- or post-condition checks immediately before or after the action description of that step. In addition, insert fallback actions after the corresponding condition check. Ensure logical ordering and clarity so the workflow can be executed step-by-step by a web agent.
\vspace{1em}

\noindent\textbf{OUTPUT REQUIREMENTS:}
\par\noindent
Present the final workflow as a numbered list of steps following the given workflow structure. For steps with condition checks and fallback actions, they must clearly label their components using the following tags:
\begin{itemize}[label=\textendash, wide, nosep, leftmargin=1.5em]
    \item \textbf{Action}: Concise description of the action.
    \item \textbf{Condition Check}: The condition to verify before or after the action.
    \item \textbf{Fallback Action}: The recovery action to take if the condition is not met.
\end{itemize}
\vspace{0.5em}

\noindent For steps with no condition checks or fallback actions, they must clearly label their components using only the action tag:
\begin{itemize}[label=\textendash, wide, nosep, leftmargin=1.5em]
    \item \textbf{Action}: Concise description of the action.
\end{itemize}
\vspace{0.5em}

\noindent The output must strictly follow the EXAMPLE OUTPUT FORMAT.
\vspace{1em}

\noindent\textbf{EXAMPLE OUTPUT FORMAT:}

\noindent For steps without condition check:

\begin{itemize}[label=\textendash, wide, nosep, leftmargin=1.5em]
    \item \textbf{Action}: Perform \llmplaceholder{Action}
\end{itemize}
\vspace{0.5em}

\noindent For steps with pre-condition check only:
\begin{itemize}[label=\textendash, wide, nosep, leftmargin=1.5em]
    \item \textbf{Condition Check}: Ensure \llmplaceholder{Condition} \quad \textbf{Fallback Action}: Retry \llmplaceholder{Fallback Action} \quad \textbf{Action}: Perform \llmplaceholder{Action} 
\end{itemize}
\vspace{0.5em}

\noindent For steps with post-condition check only:

\begin{itemize}[label=\textendash, wide, nosep, leftmargin=1.5em]
    \item \textbf{Action}: Perform \llmplaceholder{Action} \quad \textbf{Condition Check}: Ensure \llmplaceholder{Condition} \quad \textbf{Fallback Action}: Retry \llmplaceholder{Fallback Action}
\end{itemize}
\vspace{0.5em}

\noindent For steps with both pre- and post-condition checks:

\begin{itemize}[label=\textendash, wide, nosep, leftmargin=1.5em]
    \item \textbf{Condition Check}: Ensure \llmplaceholder{Condition} \quad \textbf{Fallback Action}: Retry \llmplaceholder{Fallback Action} \quad \textbf{Action}: Perform \llmplaceholder{Action} \quad \textbf{Condition Check}: Ensure \llmplaceholder{Condition} \quad \textbf{Fallback Action}: Retry \llmplaceholder{Fallback Action}
\end{itemize}

\end{llmscriptblock}

\section{Benchmark Evaluation Results} \label{sec:benchmark_eval_sr_mean_ci}
Table \ref{tab:auto_eval_srs_avgcis} presents the mean success rates (\%) with 95\% confidence intervals of each task and across tasks for each condition in the benchmark evaluation (Section \ref{sec:benchmark_evaluation}). 

\begin{table}[!ht]
\centering
\footnotesize
\caption{Benchmark evaluation results. Mean success rates (\%) $\uparrow$ with 95\% confidence intervals of each task and across tasks.}

\begin{tabular}{@{}c c@{}}
% ================= LEFT BLOCK (STACKED TASK TABLES) =================
\begin{minipage}[t]{0.78\linewidth}
\centering

% ----------- Part 1 -----------
\begin{tabular}{lp{0.18\textwidth}p{0.18\textwidth}p{0.18\textwidth}}
\toprule
 &
\textbf{Archive} &
\textbf{Asus} &
\textbf{Cars} \\
\midrule
Task-Only &
13.3 [-15.4, 42.0] &
45.0 [14.5, 75.5] &
20.0 [20.0, 20.0] \\

Task + Success-Traces &
26.7 [-2.0, 55.4] &
80.0 [80.0, 80.0] &
30.0 [-97.1, 157.1] \\

Task + Magentic-UI Plan &
33.3 [4.6, 62.0] &
50.0 [-5.1, 105.1] &
50.0 [-77.1, 177.1] \\

Task + ReUseIt Workflow &
\textbf{86.7} [58.0, 115.4] &
\textbf{90.0} [71.6, 108.4] &
\textbf{80.0} [80.0, 80.0] \\
\bottomrule
\end{tabular}

\vspace{0.6em}

% ----------- Part 2 -----------
\begin{tabular}{lp{0.18\textwidth}p{0.18\textwidth}p{0.18\textwidth}}
\toprule
 &
\textbf{Bandcamp} &
\textbf{BBC} &
\textbf{Fda} \\
\midrule
Task-Only &
40.0 [-214.1, 294.1] &
25.0 [-5.5, 55.5] &
15.0 [-15.5, 45.5] \\

Task + Success-Traces &
50.0 [-331.2, 431.2] &
45.0 [-2.7, 92.7] &
35.0 [4.5, 65.5] \\

Task + Magentic-UI Plan &
80.0 [80.0, 80.0] &
60.0 [15.0, 105.0] &
40.0 [14.0, 66.0] \\

Task + ReUseIt Workflow &
\textbf{90.0} [-37.1, 217.1] &
\textbf{80.0} [54.0, 106.0] &
\textbf{55.0} [39.1, 70.9] \\
\bottomrule
\end{tabular}

\vspace{0.6em}

% ----------- Part 3 -----------
\begin{tabular}{lp{0.18\textwidth}p{0.18\textwidth}p{0.18\textwidth}}
\toprule
 &
\textbf{Forbes} &
\textbf{Gettyimages} &
\textbf{Groupon} \\
\midrule
Task-Only &
15.0 [-32.7, 62.7] &
10.0 [-21.8, 41.8] &
40.0 [14.0, 66.0] \\

Task + Success-Traces &
30.0 [11.6, 48.4] &
30.0 [11.6, 48.4] &
60.0 [23.3, 96.7] \\

Task + Magentic-UI Plan &
35.0 [-12.7, 82.7] &
35.0 [4.5, 65.5] &
45.0 [5.0, 85.0] \\

Task + ReUseIt Workflow &
\textbf{45.0} [-2.7, 92.7] &
\textbf{55.0} [39.1, 70.9] &
\textbf{80.0} [54.0, 106.0] \\
\bottomrule
\end{tabular}

\vspace{0.6em}

% ----------- Part 4 -----------
\begin{tabular}{lp{0.18\textwidth}p{0.18\textwidth}p{0.18\textwidth}}
\toprule
 &
\textbf{Restaurantguru} &
\textbf{Scribd} &
\textbf{Smithsonianmag} \\
\midrule
Task-Only &
20.0 [-25.0, 65.0] &
10.0 [-8.4, 28.4] &
35.0 [-12.7, 82.7] \\

Task + Success-Traces &
30.0 [11.6, 48.4] &
35.0 [19.1, 50.9] &
55.0 [7.3, 102.7] \\

Task + Magentic-UI Plan &
45.0 [5.0, 85.0] &
40.0 [14.0, 66.0] &
65.0 [49.1, 80.9] \\

Task + ReUseIt Workflow &
\textbf{50.0} [-16.2, 116.2] &
\textbf{55.0} [24.5, 85.5] &
\textbf{90.0} [58.2, 121.8] \\
\bottomrule
\end{tabular}

\vspace{0.6em}

% ----------- Part 5 -----------
\begin{tabular}{lp{0.18\textwidth}p{0.18\textwidth}p{0.18\textwidth}}
\toprule
 &
\textbf{Sportskeeda} &
\textbf{Ticketmaster} &
\textbf{Tvguide} \\
\midrule
Task-Only &
20.0 [-25.0, 65.0] &
45.0 [14.5, 75.5] &
10.0 [-21.8, 41.8] \\

Task + Success-Traces &
40.0 [40.0, 40.0] &
45.0 [5.0, 85.0] &
30.0 [-1.8, 61.8] \\

Task + Magentic-UI Plan &
40.0 [-5.0, 85.0] &
55.0 [39.1, 70.9] &
55.0 [39.1, 70.9] \\

Task + ReUseIt Workflow &
\textbf{55.0} [15.0, 95.0] &
\textbf{65.0} [49.1, 80.9] &
\textbf{75.0} [27.3, 122.7] \\
\bottomrule
\end{tabular}

\end{minipage}
&
% ================= RIGHT BLOCK (ACROSS TASKS) =================
\begin{minipage}[t]{0.20\linewidth}
\centering

\begin{tabular}{lc}
\toprule
\textbf{Across Tasks} \\
\midrule
24.2 [16.9, 31.5] \\
41.4 [33.2, 49.6] \\
48.6 [41.4, 55.7] \\
\textbf{70.1} [61.0, 79.2] \\
\bottomrule
\end{tabular}

\end{minipage}

\end{tabular}
\label{tab:auto_eval_srs_avgcis}
\end{table}

\section{User Study Materials and Results} \label{appendix:user_eval_materials}
This section provides the materials (interview questions) used in the user study (Section \ref{sec:user_study_design}) and additional results of the follow-up experiments (Section \ref{sec:user_eval_followup}).

\subsection{Interview Questions} \label{appendix:user_eval_interview}
\subsubsection*{Understanding of Agent Actions and Failures}
\begin{enumerate}
    \item You have examined two types of workflows: (1) Magentic-UI Plan, and (2) ReUseIt Workflow. Can you share your insights into the two types of workflows regarding:
    \begin{enumerate}
        \item How well did each type of workflow support your understanding of the agent actions? Why do you think so?
        \item How well did the agent messages produced when guided by the two types of workflows support your understanding of the agent issues?
        \begin{enumerate}
            \item For Magentic-UI Plan, what did you find helpful and not helpful? Why do you think so?
            \item For ReUseIt Workflow, what did you find helpful and not helpful? Why do you think so?
        \end{enumerate}
    \end{enumerate}

    \item You have also examined the agent messages produced with Task-Only. Can you share your insights regarding how well these messages support your understanding of the agent issues?
    \begin{enumerate}
        \item What did you find helpful and not helpful? Why do you think so?
    \end{enumerate}
\end{enumerate}

\subsubsection*{User Guidance}
\begin{enumerate}[resume]
    \item How did the agent messages help you to provide guidance?
    \begin{enumerate}
        \item When the messages were produced based on ReUseIt Workflow. Why do you think so?
        \item When the messages were produced based on Magentic-UI Plan. Why do you think so?
        \item When the messages were produced based on Task-Only. Why do you think so?
    \end{enumerate}

    \item Do you think the workflows help you understand what guidance can be helpful?
    \begin{enumerate}
        \item If yes, can you tell me an example and expand on how the workflow helped you provide guidance?
    \end{enumerate}
\end{enumerate}

\subsubsection*{Reuse of Web Agent}
\begin{enumerate}[resume]
    \item Comparing the three workflow conditions, which condition and the corresponding LLM message do you prefer in order to reuse the web agent to help you execute the same or similar tasks?
    \begin{enumerate}
        \item Why do you think so? (Impact of agent messages and workflow on user-perceived reliability for reuse)
    \end{enumerate}

    \item If you were able to include more information or add more features, what else do you think could be helpful to boost your understanding of web agents and reliably reuse web agents? Or any suggested improvements to use agent messages and workflows?

    \item Is there anything else you’d like to share about your experience in this study?
\end{enumerate}

\subsection{Follow-Up Experiment Results} 

\subsubsection{Mean Success Rates with Confidence Intervals} \label{sec:user_study_sr_mean_ci}
Table \ref{tab:user_study_srs_avgcis} reports the mean success rates (\%) with 95\% confidence intervals of each task and across tasks for each condition in the user study. The difference between this table and Table~\ref{tab:prelim_eval} is the success rates of additional conditions---Task + ReUseIt Workflow and Task + ReUseIt Workflow User---for each task and across all tasks. 

\begin{table}[!ht]
\centering
\footnotesize
\caption{User study results. Mean success rates (\%) $\uparrow$ with 95\% confidence intervals of each task and across tasks.}

\begin{tabular}{@{}c c@{}}
% ================= LEFT BLOCK (TWO STACKED TABLES) =================
\begin{minipage}[t]{0.78\linewidth}
\centering

% ----------- Part 1: First three tasks -----------
\begin{tabular}{lp{0.18\textwidth}p{0.18\textwidth}p{0.18\textwidth}}
\toprule
 &
\textbf{Flight} &
\textbf{Housing} &
\textbf{Pet} \\
\midrule
Task-Only &
17.5 [9.5, 25.5] &
36.7 [22.3, 51.0] &
26.7 [-2.0, 55.4] \\

Task + Success-Traces &
40.0 [17.5, 62.5] &
80.0 [-6.1, 166.1] &
73.3 [16.0, 130.7] \\

Task + Magentic-UI Plan &
42.5 [22.5, 62.5] &
80.0 [30.3, 129.7] &
80.0 [30.3, 129.7] \\

Task + Magentic-UI Plan User &
45.0 [34.8, 55.2] &
82.2 [63.1, 101.3] &
84.4 [65.3, 103.6] \\

Task + ReUseIt Workflow &
67.5 [59.5, 75.5] &
86.7 [29.3, 144.0] &
86.7 [58.0, 115.4] \\

Task + ReUseIt Workflow User &
\textbf{76.7} [70.5, 82.8] &
\textbf{90.0} [90.0, 90.0] &
\textbf{88.9} [63.6, 114.2] \\
\bottomrule
\end{tabular}

\vspace{0.6em}

% ----------- Part 2: Next three tasks -----------
\begin{tabular}{lp{0.18\textwidth}p{0.18\textwidth}p{0.18\textwidth}}
\toprule
 &
\textbf{Price} &
\textbf{Publication} &
\textbf{News} \\
\midrule
Task-Only &
30.0 [11.6, 48.4] &
26.7 [-2.0, 55.4] &
30.0 [11.6, 48.4] \\

Task + Success-Traces &
45.0 [-15.2, 105.2] &
60.0 [-26.1, 146.1] &
70.0 [9.1, 130.9] \\

Task + Magentic-UI Plan &
62.5 [20.7, 104.3] &
63.3 [25.4, 101.3] &
80.0 [35.0, 125.0] \\

Task + Magentic-UI Plan User &
73.3 [50.4, 96.2] &
71.1 [36.6, 105.6] &
85.0 [58.5, 111.5] \\

Task + ReUseIt Workflow &
\textbf{90.0} [71.6, 108.4] &
\textbf{93.3} [64.6, 122.0] &
\textbf{95.0} [79.1, 110.9] \\

Task + ReUseIt Workflow User &
-- &
-- &
92.5 [84.5, 100.5] \\
\bottomrule
\end{tabular}

\end{minipage}
&
% ================= RIGHT BLOCK (ACROSS TASKS) =================
\begin{minipage}[t]{0.20\linewidth}
\centering

\begin{tabular}{lc}
\toprule
\textbf{Across Tasks} \\
\midrule
27.9 [21.3, 34.5] \\
61.4 [44.5, 78.3] \\
68.1 [52.2, 83.9] \\
73.5 [57.6, 89.4] \\
86.5 [76.1, 96.9] \\
\textbf{87.0} [75.8, 98.3] \\
\bottomrule
\end{tabular}

\end{minipage}

\end{tabular}
\label{tab:user_study_srs_avgcis}
\end{table}

\subsubsection{Per-User Results} \label{appendix:userstudy_followup_results_full}
Table \ref{tab:user_task_success_s2} shows the follow-up experiment results for each user. 

\begin{table}[!ht]
\centering
\footnotesize
\caption{User study follow-up experiment per-user results. Mean success rates (\%) $\pm$ std.}
\begin{tabularx}{0.98\linewidth}{@{}c | *{6}{Y} | l@{}}
\toprule
 & \textbf{Flight} & \textbf{Housing} & \textbf{Pet} & \textbf{Price} & \textbf{Publication} & \textbf{News} & \textbf{Across Tasks} \\
\hline
\textbf{S2-P1} & -- & 93.3 $\pm$ 11.5 & 86.7 $\pm$ 11.5 & -- & -- & -- & {90.0 $\pm$ 4.7} \\
\textbf{S2-P2} & 80.0 $\pm$ 0.0 & -- & -- & -- & -- & -- & {80.0 $\pm$ 0.0} \\
\textbf{S2-P3} & 80.0 $\pm$ 0.0 & -- & -- & -- & -- & -- & {80.0 $\pm$ 0.0} \\
\textbf{S2-P4} & -- & 86.7 $\pm$ 11.5 & 93.3 $\pm$ 11.5 & -- & -- & -- & {90.0 $\pm$ 4.7} \\
\textbf{S2-P5} & -- & -- & 86.7 $\pm$ 11.5 & -- & -- & -- & {86.7 $\pm$ 0.0} \\
\textbf{S2-P6} & -- & -- & -- & -- & -- & -- & {--} \\
\textbf{S2-P7} & -- & -- & -- & -- & -- & 90.0 $\pm$ 11.5 & {90.0 $\pm$ 0.0} \\
\textbf{S2-P8} & 70.0 $\pm$ 11.5 & -- & -- & -- & -- & -- & {70.0 $\pm$ 0.0} \\
\textbf{S2-P9} & -- & -- & -- & -- & -- & 95.0 $\pm$ 10.0 & {95.0 $\pm$ 0.0} \\
\bottomrule
\end{tabularx}
\label{tab:user_task_success_s2}
\end{table}